\documentclass[a4paper,11pt]{article}
\usepackage{jheppub}
\usepackage{lineno}
\usepackage{tikz-feynman}
\usepackage{slashed}
\usepackage[T1]{fontenc} 
\usepackage[utf8]{inputenc}
\usepackage{pdflscape}
\usepackage{amssymb}
\usepackage{tikz}
\usepackage{subfigure}
\usepackage{multirow}
\usepackage{physics}
\usepackage{makecell}
\usepackage{subcaption}
\usepackage{graphicx}
\usepackage{booktabs}
\usepackage{bm}
\usepackage{tabularx}
\usepackage{orcidlink}
\usepackage[normalem]{ulem} 
\usepackage{tabularray}
\makeatletter
\gdef\@fpheader{}
\makeatother

\newcommand{\bpm}{\begin{pmatrix}}      
\newcommand{\epm}{\end{pmatrix}}

\definecolor{lightgray}{rgb}{0.83, 0.83, 0.83}
\definecolor{lightpurp}{rgb}{0.901,0.796,0.882}

\newcommand{\VR}[1]{\textcolor{green}{#1}}

\title{\boldmath Testing the Spin of New Physics with Top Precision Measurements}

\author[a]{Andr\'e Lessa\orcidlink{0000-0002-5251-7891}}
\author[a]{Vin\'icius Rocca\orcidlink{0000-0001-8249-8014}}
\author[b]{and Veronica Sanz\orcidlink{0000-0001-8864-2507}}
\affiliation[a]{Instituto de Física, Universidade de São Paulo,
São Paulo, 05508-090 SP, Brazil}
\affiliation[b]{Instituto de F\'isica Corpuscular (IFIC), Universidad de Valencia-CSIC, E-46980 Valencia, Spain}

\emailAdd{lessa@if.usp.br}
\emailAdd{rocca@if.usp.br}
\emailAdd{veronica.sanz@uv.es}

\abstract{
Precision measurements can reveal not only the presence of new physics, but also information about the dynamics that generated it. We study this question in top-pair production by comparing three top-philic scenarios: a colored scalar top partner, a colored vector-like fermion, and a spin-one $Z^\prime$ mediator. The first two contribute through loop corrections, while the $Z^\prime$ contributes at tree level. We show that the scalar and vector-like fermion scenarios match onto the same effective operator structure at low energies,
 making their distinction unfeasible using the EFT approach. However, near or above threshold their full loop amplitudes develop distinct momentum-dependent interference patterns. Using the top transverse momentum spectrum we find that resonant and loop-induced effects are relatively easy to distinguish, whereas separating the scalar and vector-like hypotheses is substantially more challenging. Nevertheless, their different analytic structures leave measurable information in the shape of the distribution.
These results illustrate how precision differential measurements can probe the spin and dynamics of virtual new states even when they are not directly reconstructed, and motivate more powerful multidimensional analyses.
 }

\begin{document}
\maketitle
\flushbottom

\section{Introduction}
\label{sec:intro}

The absence of unambiguous signals of new physics at the LHC has increased the importance of
precision measurements as probes of physics beyond the Standard Model (BSM).  This shift
is particularly consequential in the top-quark sector.  Owing to its large Yukawa coupling
and its proximity to the electroweak scale, the top quark is a natural portal to dynamics
connected with electroweak symmetry breaking, naturalness, and dark matter.  At the same
time, the large Run~2 and Run~3 samples and the luminosity expected at the High-Luminosity LHC
(HL-LHC) turn differential top measurements into precision observables over a wide range
of energies~\cite{Azzi:2019yne}.  Measurements of $t\bar t$ production now resolve the
invariant mass, transverse momentum, rapidity, and additional-radiation dependence in
several decay channels~\cite{CMS-TOP-20-001,ATLAS-TOPQ-2019-23,ATLAS:2023gsl,CMS:2024ybg} as well as the $t\bar{t}$ spin correlation~\cite{CMS:2024zkc}.
Their interpretation is supported by fully differential QCD predictions at
next-to-next-to-leading order~\cite{Catani:2019iny,Catani:2019hip}.  Consequently, a small
but coherent departure from the Standard Model (SM) need not first appear as a clear resonance: it
may instead emerge as a structured deformation of a precisely measured spectrum.

The Standard Model Effective Field Theory (SMEFT) provides the systematic framework
most commonly used to organize such deviations.  Once heavy new degrees of freedom are
integrated out, their leading effects are encoded in higher-dimensional operators built
from SM fields~\cite{Grzadkowski:2010es,Brivio:2017vri}.  Differential top data have become
an essential ingredient of both dedicated top-sector analyses and global SMEFT
fits~\cite{Degrande:2010kt,Buckley:2015lku,Hartland:2019bjb,Brivio:2019ius,Ellis:2020unq,Kassabov:2023hbm}.
This description is especially powerful when the characteristic momentum transfer is
well below the masses of the new particles: a relatively small set of Wilson coefficients
then captures the leading low-energy imprint of a broad class of ultraviolet (UV)
theories, and can in turn be related to discoverable UV scenarios
~\cite{Cepedello:2022pyx}.

The same expansion also defines its own limitation.  If the particles responsible for a
deviation are not parametrically heavier than the energies probed, the expansion in
external momenta over the new-physics scale need not converge rapidly.  Thresholds,
absorptive parts, and changes of sign are non-local features of amplitudes and cannot, in
general, be reproduced by a truncation at a fixed EFT dimension.  Exact one-loop and EFT
descriptions can therefore lead to significantly different conclusions even when they arise
from the same UV model~\cite{Drozd:2015kva}.  In Ref.~\cite{Lessa_2024}, we developed
this observation in the context of top-pair production.  For a scalar top partner and a
singlet fermion protected by a $\mathcal Z_2$ symmetry, we showed that the full
momentum-dependent form factors generate characteristic structures in the $m_{t\bar t}$
and top-quark transverse-momentum spectra.  These structures may be largest around the
new-particle threshold and decrease again in the far tail, in contrast with the monotonic
energy growth often associated with a dimension-six interpretation.

This raises a question that is logically subsequent to exclusion or discovery: if a
deformation is observed, how much can its shape reveal about the dynamics that produced
it?  Determining the quantum numbers of new states has long been recognized as a central
part of the LHC programme.  Most spin studies exploit on-shell production, decay-angle
correlations, endpoints, or production rates~\cite{Meade:2006dw,Wang:2008sw}.  For top
partners followed by invisible particles, direct $t\bar t+E_T^{\rm miss}$ searches and
their kinematic distributions likewise carry information about both the spectrum and the
spin assignment~\cite{Boughezal:2013pja}.  Such searches are indispensable, but they can
lose sensitivity for compressed spectra~\cite{CMS-SUS-20-002}, while the same particles can
still leave virtual effects in high-statistics SM measurements.  This motivates asking
whether spin-dependent information can be recovered indirectly from precision spectra,
without requiring the new states to be reconstructed on shell.

Top-philic sectors provide a well-motivated setting for this question.  Scalar and
fermionic colored partners coupled to a neutral stable state occur in simplified dark
matter models~\cite{Abercrombie:2015wmb,Arina:2016cqj,Garny:2018icg,Colucci_2018,Arina:2025zpi}.
Spin-one
top-philic mediators offer a qualitatively different realization and have been studied in
collider and dark-matter phenomenology~\cite{Cox:2015afa,Fox:2018ldq}.  The experimental
program is now directly probing related structures, including heavy scalar and
pseudoscalar states in $t\bar t$~\cite{CMS:2025dzq} and top-philic $Z'$ bosons at
$\sqrt{s}=13.6$~TeV~\cite{CMS:2026gdz}.  These examples also illustrate why the
$m_{t\bar t}$ spectrum is especially informative: it can resolve a pole and its
interference pattern, but also the threshold and dispersive structure generated by new
particles contributing at one loop.

In this work, we investigate whether these qualitatively different dynamical origins can
be distinguished using the shape of $t\bar t$ distributions.  We compare
three representative hypotheses: a colored scalar top partner coupled to a singlet
fermion, a colored vector-like fermion coupled to a singlet scalar, and a top-philic
$Z'$ boson.  The first two modify $t\bar t$ production at one loop, whereas the third
contributes through an $s$-channel amplitude already at tree level.  The comparison is
therefore not intended as a model-independent extraction of spin with all other physical
properties held fixed.  Rather, it tests whether representative spin-zero, spin-one-half,
and spin-one top-philic hypotheses leave distinguishable analytic and kinematic structures
once their masses and signal strengths are allowed to vary.

We first illustrate the different analytic structures generated at one loop by the spin-zero and spin-one-half models through a discussion of their renormalized top--gluon form factors and their impact on the $t\bar{t}$ distributions.  
The $p_T$ distribution is then employed to quantify the distinguishability of each signal hypothesis under the assumption of an observed excess. 
We first determine the relevant model parameters of each individual hypothesis which best fit the excess and then employ a binned profile-likelihood analysis to
quantify the luminosity required to reject the wrong signal hypothesis.

The paper is organized as follows.  In Sec.~\ref{sec:beyondtop} we introduce the three
top-philic scenarios.  Section~\ref{sec:observables} presents their contributions to
$t\bar t$ production, including the form-factor analysis and differential distributions.
Section~\ref{sec:directSearches} gives a brief overview of the constraints from direct searches for top partners. 
In Sec.~\ref{sec:shape_discrimination} we construct the shape-discrimination test and present
the expected sensitivity.  We summarize our conclusions in
Sec.~\ref{sec:conclusions}.

\section{Top-Philic BSM Scenarios}\label{sec:beyondtop}

We consider three simple UV extensions of the SM covering distinct spin assignments for the BSM states:
\begin{enumerate}
    \item {\bf Scalar Top-partner}:  contains a top-like spin $0$ particle singlet under $SU(2)_L$, $\varphi_T$, and a spin $1/2$ particle, $\chi$, which is a singlet under the entire SM gauge group and plays the role of a DM candidate. Both $\varphi_T$ and $\chi$ are odd under a $\mathcal{Z}_2$-type symmetry.
    \item {\bf Vector-like Fermionic Top-partner} (VLF): contains a vector-like top-partner, $\psi_T$, which is also a singlet under $SU(2)_L$ and a fermionic field, $\phi$, which is a DM candidate.  Both $\psi_T$ and $\phi$ are odd under a $\mathcal{Z}_2$-type symmetry.
    \item {\bf Top-philic $Z'$}: correspond to a simplified $U(1)'$ model with couplings to light quarks and top quarks. We assume, however, that the couplings to top quarks are enhanced, so $Z'$ mostly decays to top pairs.\footnote{Note that although we could also include a DM candidate for the $Z^{\prime}$ scenario, it will have no significant impact on the top observables discussed here.}
\end{enumerate}

Using the particle content for the three scenarios described above and assuming top-philic couplings we can write the following Lagrangians for the new fields:
\begin{flalign}
    \mathcal{L}_{\mathrm{Scalar}} & =
\bar{\chi}\left( i \slashed{\partial} - \frac{1}{2} m_{\chi} \right) \chi
+ |D_\mu \varphi_T|^2
- m_{\varphi}^2 |\varphi_T|^2
- \left( y_{\mathrm{DM}}\, \varphi_T^\dagger\, \bar{\chi} \, t_R + \text{h.c.} \right) \label{eq:Lscalar}\\
\mathcal{L}_{\mathrm{VLF}} & =
\bar{\psi}_T \left( i \gamma^\mu D_\mu - m_{\psi} \right) \psi_T
+ \frac{1}{2}(\partial_\mu \phi)^2
- \frac{1}{2} m_{\phi}^2 \phi^2
- \left( y_{\mathrm{DM}}\, \bar{\psi}_T \, t_R\, \phi + \text{h.c.} \right) \label{eq:LVLF}\\
    \mathcal{L}_{Z^\prime} & = 
    -\frac{1}{4} Z^\prime_{\mu\nu} Z^{\prime\mu\nu}
    - \frac{1}{2} m_{Z^\prime}^2 Z^\prime_\mu Z^{\prime\mu} 
    - g_q  Z^{\prime\mu} \sum_{i=1}^{2}\left(\bar{u}_i \gamma_\mu  u_i +\bar{d}_i \gamma_\mu  d_i\right)
    - g_t  Z^{\prime\mu} \bar{t} \gamma_\mu  t 
    \label{eq:LZprime}
\end{flalign}
The models presented above are simply intended to parametrize the effects of distinct BSM particle content on top observables. Therefore we do not impose additional experimental constrains on the model parameters except for limits from direct searches.

For the Scalar and VLF scenarios we always assume the DM candidate to be lighter than the top partner, so the latter always decays to the DM and a top. Depending on the mass difference,  $\Delta M = m_\varphi - m_{\chi}$ (or $\Delta M = m_{\psi} - m_{\phi}$) the decay can be through an off-shell top. In these scenarios the interaction between the dark matter particle and the SM are entirely controlled by the coupling $y_{\mathrm{DM}}$.
Although the Scalar scenario shares some phenomenological similarities with the Minimal Supersymmetric Standard Model (MSSM) featuring a Bino lightest supersymmetric particle and a right-handed stop, an important difference should be emphasized. In the supersymmetric case, the corresponding coupling is determined by the composition of the lightest supersymmetric particle and is typically of the order of the electroweak couplings ($y_{\mathrm{DM}} \sim 0.1$). In contrast, in the framework considered here $y_{\mathrm{DM}}$ is treated as a free parameter and can take values as large as allowed by perturbativity, $y_{\mathrm{DM}} \lesssim 4\pi$.

The $Z^\prime$ model can also accommodate a Dark Matter candidate and is a standard benchmark for resonance and dark matter searches at the LHC, as extensively discussed in the literature \cite{Abercrombie:2015wmb, Boveia:2018yeb, okada2018zprimeportaldarkmatter, wang2025revivingzprimeportaldark, ghosh2025darkmatterfreezeinzprime}.
Particularly relevant for our study is the coupling to top-quark $g_t$, which controls the resonant $s$-channel production of top-quark pairs. As in the previous models, we treat the $Z^\prime$ mass, its total width, and its couplings as free parameters. This approach allows us to explore the distinct kinematic features that may be imprinted on differential top-quark observables while remaining agnostic about the precise BSM model.

The dark matter phenomenology of the three scenarios has been extensively studied in literature, e.g. Ref.~\cite{Arina:2025zpi,Garny:2018icg,Arina:2016cqj,Baek_2016,Colucci_2018,Abercrombie:2015wmb, Boveia:2018yeb, okada2018zprimeportaldarkmatter,ghosh2025darkmatterfreezeinzprime,wang2025revivingzprimeportaldark}.
In all scenarios the observed relic abundance can be obtained over a broad range of masses and couplings. In particular, relatively large couplings, $y_{\mathrm{DM}} \gtrsim 3$, are required in certain regions of parameter space. In the present work, however, we do not impose dark matter constraints, since these may be significantly altered by the presence of additional heavy BSM states or by modifications to the standard cosmological history.

The models introduced above can lead to implications for LHC phenomenology and precision SM observables. In this work, however, we focus primarily on the indirect effects of the BSM particles in precision measurements involving top quarks.
Despite being briefly discussed in Sec.~\ref{sec:directSearches},  a comprehensive study of direct and indirect constraints on these BSM scenarios is left for future work.

\section{Impact on Top Observables}\label{Sec:Impact_Top_observables}
\label{sec:observables}
The BSM models considered here couple predominantly to the top quark and can modify both the total production rate of top-quark pairs and the shapes of their kinematic distributions. Focusing on the process
$pp\to t\bar t$, we compare the leading BSM effects in the three scenarios introduced in Sec.~\ref{sec:beyondtop}. In the Scalar and VLF models, the leading corrections arise at one loop from the new BSM states, whereas in the $Z^\prime$ model the leading contribution already appears at tree level through $s$-channel $Z^\prime$ exchange.

\subsection{BSM Form Factors}\label{sec:formFactors}

\begin{figure}[htpb]
    \centering
    \begin{minipage}[b]{0.28\textwidth}
        \centering
        \includegraphics[scale=0.95]{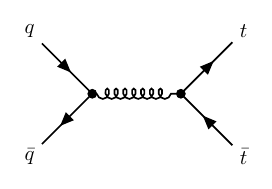} \\
        (a)
    \end{minipage}
    \hfill
    \begin{minipage}[b]{0.33\textwidth}
        \centering
        \includegraphics[scale=0.95]{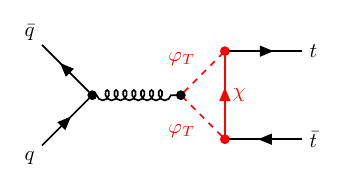}\\
        (b)
    \end{minipage}
    \hfill
    \hspace{7pt}
    \begin{minipage}[b]{0.33\textwidth}
        \centering
        \includegraphics[scale=0.95]{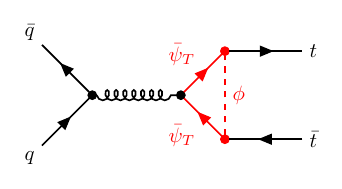}\\
        (c)
    \end{minipage}
    \caption{
    Representative Feynman diagrams for the
    $q\bar q\to t\bar t$ process:
    (a) tree-level SM contribution;
    (b) leading one-loop correction in the Scalar model;
    and (c) leading one-loop correction in the VLF model.
    }
    \label{fig:production_diagrams}
\end{figure}

Before presenting the complete one-loop predictions, it is useful to isolate the structure of the loop corrections generated by the Scalar and VLF models. For this purpose, we first restrict the discussion to the quark-initiated subprocess,
$q\bar q\to t\bar t$, for which the relevant BSM contribution can be traced back to the loop-induced correction to the $g$-$t$-$\bar t$ vertex, as shown in Fig.~\ref{fig:production_diagrams}.

We write this effective vertex as
\begin{equation}
    \mathcal{L}_{i} =
    \pi^2 g_s y_{DM}^2 \,
    G_\mu^A\,
    \bar t
    \left[
    T^A \mathcal{F}_{i}^{\mu}(p_t,p_{\bar t})
    \right]t,
    \qquad
    i=\mathrm{Scalar,VLF},
\label{eq:formFactors}
\end{equation}
where $T^A$ are the Gell-Mann matrices and the form factor $\mathcal{F}_{i}^{\mu}$ contains both the momentum dependence and the Lorentz structure induced by the loop.

Using $M$ to denote generically the mass of the colored top partner and $m$ the mass of the DM candidate, the renormalized form factors can be written as
\begin{align}
    \mathcal{F}^\mu_{\mathrm{Scalar}}(p_1,p_2)
    &=
    \frac{1}{\pi^2}\int \frac{d^D q}{(2\pi)^D}
    \frac{
    (\slashed q+\slashed p_1)
    (p_1^\mu+p_2^\mu+2q^\mu)
    }
    {[(q+p_1)^2-m^2][(q+p_1+p_2)^2-M^2](q^2-M^2)}
    P_R
    \nonumber\\
    &\quad
    +2\gamma^\mu
    \left(
    \delta_L P_L+\delta_R P_R
    \right),
\label{eq:FscalarAmp}
\\[1.5ex]
    \mathcal{F}^\mu_{\mathrm{VLF}}(p_1,p_2)
    &=
    -\frac{1}{\pi^2}\int \frac{d^D q}{(2\pi)^D}
    \frac{
    M^2\gamma^\mu
    +(\slashed p_1+\slashed p_2+\slashed q)
    \gamma^\mu\slashed q
    }
    {[(q+p_1)^2-m^2][(q+p_1+p_2)^2-M^2](q^2-M^2)}
    P_R
    \nonumber\\
    &\quad
    +2\gamma^\mu
    \left(
    \delta_L P_L+\delta_R P_R
    \right). 
\label{eq:FVLFamp}
\end{align}
Here $\delta_{L,R}$ denote the corresponding counterterms required to renormalize the top-quark self-energy and the $t$-$\bar t$-$g$ vertex, while
$P_{R,L}=(1\pm\gamma^5)/2$ are the chiral projectors.
The detailed expressions for the VLF and Scalar counterterms can be found in Appendix~\ref{app:CTs}.

The loop integrals were evaluated using
\texttt{FeynArts}~\cite{Hahn:2000kx} and
\texttt{FeynCalc}~\cite{Shtabovenko:2020gxv,Shtabovenko:2016sxi,Mertig:1990an},
while the counterterms were obtained with
\texttt{NLOCT}~\cite{Degrande:2014vpa}
in the on-shell renormalization scheme.
After Passarino--Veltman reduction~\cite{Passarino:1978jh,tHooft:1978jhc}, the form factors take the form
\begin{align}
    \mathcal{F}_{\mathrm{Scalar}}^\mu
    &=
    \slashed p_1 P_R
    \left[
    p_1^\mu(C_1+2C_{11})
    -
    p_2^\mu(C_1+2C_{12})
    \right]
    \nonumber\\
    &\quad
    +
    \slashed p_2 P_R
    \left[
    p_2^\mu(C_2+2C_{22})
    -
    p_1^\mu(C_2+2C_{12})
    \right]
    \nonumber\\
    &\quad
    +
    2\gamma^\mu
    (C_{00}+\delta_R)P_R
    +
    2\gamma^\mu\delta_L P_L ,
\label{eq:fmu}
\\[1ex]
    \mathcal{F}_{\mathrm{VLF}}^\mu
    &=
    -\slashed p_1 P_R
    \left[
    p_1^\mu(2C_1+2C_{11})
    -
    2p_2^\mu C_{12}
    \right]
    \nonumber\\
    &\quad
    -
    \slashed p_2 P_R
    \left[
    p_2^\mu(2C_2+2C_{22})
    -
    2p_1^\mu C_{12}
    \right]
    \nonumber\\
    &\quad
    +
    \slashed p_2\gamma^\mu\slashed p_1 P_R
    (C_0+C_1+C_2)
    \nonumber\\
    &\quad
    +
    \gamma^\mu P_R
    \left[
    B_0+(m^2-M^2)C_0
    +2p_1^2C_1+2p_2^2C_2
    \right]
    \nonumber\\
    &\quad
    -
    2\gamma^\mu P_R(C_{00}-\delta_R)
    +
    2\gamma^\mu\delta_L P_L .
\label{eq:FmuVLF}
\end{align}
Here the three-point functions have arguments
\begin{equation}
    C_i =
    C_i\!\left(
    p_1^2,(p_1+p_2)^2,p_2^2;
    m^2,M^2,M^2
    \right),
\end{equation}
with analogous conventions for $C_{ij}$ and $C_{00}$, while the corresponding arguments of $B_0$ are implicit.

The different structure of Eqs.~\eqref{eq:fmu} and \eqref{eq:FmuVLF} originates directly from the spin assignments of the particles running in the loop. In the Scalar model, the triangle contains one fermion and two scalar propagators, whereas in the VLF model it contains two fermion propagators and one scalar propagator. The additional momentum dependence in the numerators of the fermionic propagators generates Lorentz structures that are absent in the Scalar model. A particularly transparent example is the
$\slashed p_2\gamma^\mu\slashed p_1$
term in Eq.~\eqref{eq:FmuVLF}.

This difference is important because it provides the microscopic origin of the distinct momentum dependence of the two models. As we show below, in the low-energy regime it appears as a different matching coefficient of the same effective interaction, while close to the production threshold it leads to genuinely different non-local loop structures.

For a more transparent analytical comparison, we now take the limit of negligible top mass,
$m_t\ll M,\sqrt{\hat{s}}$, where $\sqrt{\hat{s}}$ is the partonic center of mass energy.
The BSM contribution to
$q\bar q\to t\bar t$
is proportional to
$\bar u_t\mathcal{F}_i^\mu v_{\bar t}$.
Using the on-shell Dirac equations for the external top spinors, the form factors simplify considerably in the $m_t\to0$ limit. If, in addition, the two BSM masses are approximately degenerate,
$m\simeq M$, the loop functions can be expressed in a compact analytical form:
\begin{equation}
    \bar u_t\mathcal{F}_{i}^{\mu}v_{\bar t}
    \simeq
    \,\frac{1}{32\pi^2}
    F_i(\hat{s}/M^2)\,
    \bar u_t\gamma^\mu P_R v_{\bar t},
    \qquad
    m_t\to0,\quad m\to M.
\label{eq:ff_simplified}
\end{equation}
The two dimensionless functions are
\begin{align}
    F_{\mathrm{Scalar}}(x)
    &=
    \frac{L(x)}{x}\left[ \Delta(x) + L(x)\right] +3,
\label{eq:fmuZero2}
\\
    F_{\mathrm{VLF}}(x)
    &=
    \frac{L(x)}{x}\left[ \Delta(x) - L(x)\right]  +1,
\label{eq:fmuZero2VLF}
\end{align}
where
\begin{align}
    L(x)  &= \log\left[1 + \frac{\Delta(x)}{2} - \frac{x}{2} \right],
\label{eq:Lfunction}
\\
    \Delta(x)
    &=
    \sqrt{x^2-4x}.
\label{eq:Deltadef}
\end{align}
The functions $F_{\mathrm{Scalar}}$ and $F_{\mathrm{VLF}}$ develop imaginary parts above the pair-production threshold,
$\hat{s} > 4M^2$, while below threshold they are real.

For both the Scalar and VLF models, we compute the squared amplitude at leading non-trivial order in $\alpha_s$ and $y_{DM}$. The SM contribution is of order
$\mathcal{O}(\alpha_s^2)$,
while the leading BSM correction enters through the interference term at
$\mathcal{O}(\alpha_s^2 y_{DM}^2)$:
\begin{equation}
    |\mathcal{M}|^2
    =
    |\mathcal{M}_{\rm QCD}|^2
    +
    2\,\mathrm{Re}
    \left(
    \mathcal{M}_{\rm QCD}^*
    \mathcal{M}_{\rm BSM}
    \right).
\label{eq:amp}
\end{equation}

In terms of Feynman diagrams, the leading SM amplitude is given by Fig.~\ref{fig:production_diagrams}(a), while the one-loop BSM corrections proportional to
$g_s^2y_{DM}^2$
are represented in
Figs.~\ref{fig:production_diagrams}(b) and
\ref{fig:production_diagrams}(c)
for the Scalar and VLF models, respectively.
The corresponding amplitudes are
\begin{align}
    \mathcal{M}_{\mathrm{QCD}}
    &=
    -\frac{i g_s^2}{\hat{s}}
    \left[
    \bar v_{\bar q}T^A\gamma^\mu u_q
    \right]
    \left[
    \bar u_t T^A\gamma_\mu v_{\bar t}
    \right],
\label{eq:amp_sm}
\\
    \mathcal{M}_{\mathrm{BSM}}
    &=
    -\frac{i g_s^2}{\hat{s}}
    y_{\mathrm{DM}}^2
    \left[
    \bar v_{\bar q}T^A\gamma_\mu u_q
    \right]
    \left[
    \bar u_t T^A
    \mathcal{F}_i^\mu
    v_{\bar t}
    \right]
    \nonumber\\
    &\simeq
    -\frac{i g_s^2}{\hat{s}}
    \frac{y_{\mathrm{DM}}^2}{32\pi^2}
    F_i(\hat{s}/M^2)
    \left[
    \bar v_{\bar q}T^A\gamma_\mu u_q
    \right]
    \left[
    \bar u_t T^A\gamma^\mu P_R v_{\bar t}
    \right],
\label{eq:amp_bsm_simplified}
\end{align}
where the second expression is valid only in the limits
$m_t\to0$ and $m\to M$.

With the phase conventions used here, the Born SM amplitude is purely imaginary. The interference term therefore depends on the real part of the loop form factor:
\begin{equation}
    \mathrm{Re}
    \left(
    \mathcal{M}_{\mathrm{QCD}}^*
    \mathcal{M}_{\mathrm{BSM}}
    \right)
    \propto
    \frac{1}{\hat{s}^2}
    \mathrm{Re}
    \left[
    F_i(\hat{s}/M^2)
    \right],
\label{eq:int}
\end{equation}
where the last relation again refers to the simplified
$m_t\to0$, $m\to M$ limit. Consequently, changes in the sign of
$\mathrm{Re}[F_i]$
translate directly into changes in the sign of the BSM interference contribution. This connection will be useful below when interpreting the invariant-mass distributions.

\begin{figure}[htpb]
    \centering
    \begin{minipage}[b]{0.32\textwidth}
        \centering
        \includegraphics[scale = 0.95]{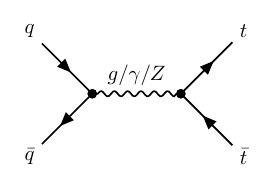}\\
        (a)
    \end{minipage}
    \begin{minipage}[b]{0.32\textwidth}
        \centering
        \includegraphics[scale = 0.95]{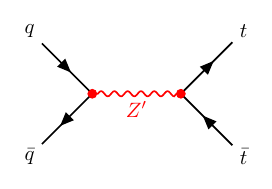} \\
        (b)
    \end{minipage}
    \caption{Relevant Feynman diagrams for $qq \to t\bar{t}$ production in the $Z^\prime$ model.}
    \label{fig:production_diagram_Zp}
\end{figure}

Before discussing the behavior of $F_{\mathrm{Scalar}}$ and $F_{\mathrm{VLF}}$ as a function of the parton level center of mass energy, it is useful to express the BSM contribution in the $Z^\prime$ model in a similar form. As discussed above, the leading BSM contribution to top quark pair production in this case arises from the tree level $s$ channel exchange of the $Z^\prime$ boson, as illustrated in Fig.~\ref{fig:production_diagram_Zp}(b). At energies well below the resonance, $\sqrt{\hat{s}}\ll m_{Z^\prime}$, and in the high energy regime, $\sqrt{\hat{s}}\gg m_{Z^\prime}$, the interference between the Born QED amplitude and the tree level $Z^\prime$ exchange can provide the dominant BSM contribution. In the resonance region, on the other hand, the squared amplitude of the tree level $Z^\prime$ exchange gives the dominant contribution. We therefore retain the following leading terms in the squared amplitude:\footnote{Note that the interference term with QCD is zero. For $\sqrt{\hat{s}} \simeq m_{Z^\prime}$ the interference terms are negligible and the squared amplitude can be approximated by $|\mathcal{M}|^2 \simeq |\mathcal{M}_{\rm QCD}|^2 + |\mathcal{M}_{\rm BSM}|^2$. However, for $\sqrt{\hat{s}} \ll m_{Z^\prime}$ or $\sqrt{\hat{s}} \gg m_{Z^\prime}$ the interference with QED  can be relevant.}
\begin{align}
    |\mathcal{M}|^2 = & |\mathcal{M}_{\rm QCD} + \mathcal{M}_{\rm QED} + \mathcal{M}_{\rm BSM}|^2 \nonumber\\
    \simeq & |\mathcal{M}_{\rm QCD}|^2 +  2\,\mathrm{Re}
    \left(
    \mathcal{M}_{\rm QED}^*
    \mathcal{M}_{\rm BSM}
    \right) + |\mathcal{M}_{\rm BSM}|^2,
\label{eq:ampZ}
\end{align}
where the BSM and QED born amplitudes are simply:
\begin{align}
\mathcal{M}_{\mathrm{BSM}}
&=  -\frac{i g_q g_t}{\hat{s} - m_{Z^\prime}^2 + i m_{Z^\prime} \Gamma_{Z^\prime}}
\left[ \bar{v}_{\bar{q}} \gamma_\mu u_q \right]
\left[ \bar{u}_t \gamma_\mu v_{\bar{t}} \right]
\label{eq:amp_Zp}\\
\mathcal{M}_{\mathrm{QED}}
&= -\frac{i e^2 Q_q Q_t}{\hat{s}}
 \left[
    \bar v_{\bar q}\gamma^\mu u_q
    \right]
    \left[
    \bar u_t \gamma_\mu v_{\bar t}
    \right]
\label{eq:amp_QED}
\end{align}
In Eq.\ref{eq:amp_Zp} we have assumed massless light quarks, $\Gamma_{Z^\prime}$ is the $Z^\prime$ width and the couplings to light quarks ($g_q$) and the top ($g_t$) are defined in Eq.\ref{eq:LZprime}, while $Q_q$ and $Q_t$ are the charge of the initial quarks and the final top quarks.

Therefore we can write the leading  $Z^\prime$ contribution to top pair production as:
\begin{align}
\left|\mathcal{M}_{\mathrm{BSM}}\right|^2 + 2\mathrm{Re}
    \left(
    \mathcal{M}_{\rm QED}^*
    \mathcal{M}_{\rm BSM}
    \right)
\propto \frac{1}{\hat{s}^2} F_{Z^\prime}(\hat{s}/m_{Z^\prime}^2)
\label{eq:amp_sq_Zp}
\end{align}
with:
\begin{align}
F_{Z^\prime}(x) \equiv 
\frac{x^2}{(1 - x)^2 + \Gamma_{Z^\prime}^2/m_{Z^\prime}^2} \left[ 1 + \frac{4Q_q}{3g_q g_t}\left(1 - \frac{1}{x}\right) \right] \label{eq:ZprimeFF}
\end{align}
Here the couplings dependence are only relevant for the region where the QED interference dominates, i.e $\sqrt{\hat{s}} \ll m_{Z^\prime}$ or $\sqrt{\hat{s}} \gg m_{Z^\prime}$.
From Eq.~\ref{eq:ZprimeFF} we see that the form factor is always real, unlike the Scalar or VLF case. 
We also point out that, for sufficiently small $x$, $F_{Z^\prime}$ can become negative.

\subsubsection*{Low Energy (EFT) Regime}

It is particularly instructive to consider the behavior of the form factors for the 3 spin choices in the limit of very heavy BSM masses and small $\Gamma_{Z^\prime}$:
\begin{equation}
    \frac{\hat{s}}{m_{Z^\prime}^2}, \frac{\hat{s}}{M^2} \ll 1 \mbox{ and } \frac{\Gamma_{Z^\prime}}{m_{Z^\prime}} \ll 1,
\end{equation}
where the particles circulating in the loop cannot be resolved and their effects admit a local effective-field-theory expansion. In this regime the functions defined in Eqs.~\ref{eq:Lfunction} and \ref{eq:Deltadef} can be written as
\begin{equation}
    L(x) =   
    2i\arcsin\left(\frac{\sqrt{x}}{2}\right)
\end{equation}
and its small-$x$ expansion gives
\begin{align}
    \frac{1}{x}L(x)^2
    &=
    -1-\frac{x}{12}
    -\frac{x^2}{90}
    +\mathcal{O}(x^3),
\\
    \frac{1}{x} \Delta(x) L(x)
    &=
    -2+\frac{x}{6}
    +\frac{x^2}{60}
    +\mathcal{O}(x^3).
\end{align}
Substitution into Eqs.~\eqref{eq:fmuZero2} and
\eqref{eq:fmuZero2VLF} yields
\begin{align}
    F_{\mathrm{Scalar}}(x)
    &=
    \frac{x}{12}
    +
    \frac{x^2}{180}
    +
    \mathcal{O}\!\left(x^3\right),
\label{eq:EFTscalar}
\\
    F_{\mathrm{VLF}}(x)
    &= \frac{x}{4}
    +\frac{x^2}{36}
    +\mathcal{O}(x^3)
\label{eq:EFTvlf}
\end{align}
while the expansion for the $Z^\prime$ form factor gives simply
\begin{equation}
    F_{Z^\prime}(x) = -\frac{4e^2 Q_q}{3g_q g_t}x + \left(1 -\frac{4e^2Q_q}{3g_q g_t}\right)x^2+ \mathcal{O}\!\left(x^3\right).
\end{equation}
Several features of these expressions are worth emphasizing. First, the momentum-independent contributions cancel exactly:
\begin{equation}
    F_{\mathrm{Scalar}}(0)
    =
    F_{\mathrm{VLF}}(0)
    =
    F_{Z^\prime}(0)
    =
    0.
\end{equation}
Indeed, the constant terms in
$F_{\mathrm{Scalar}}$ combine as $-1-2+3=0$,
while those in $F_{\mathrm{VLF}}$ combine as
$+1-2+1=0$.
This cancellation ensures that the renormalized gluon coupling matches the SM value at zero momentum.
The leading physical effect therefore decouples as $x = \hat{s}/M^2$, as expected for a dimension-six interaction.

Second, the VLF and Scalar models generate different coefficients for the leading contribution:
\begin{equation}
    F_{\mathrm{VLF}}
    =
    3F_{\mathrm{Scalar}}
    +
    \mathcal{O}\!\left(\frac{\hat{s}^2}{M^4}\right) 
\label{eq:EFTrelation}
\end{equation}
The factor of three arises from the different spin and numerator structures of the particles running in the loop. The $Z^\prime$ contribution also has a different leading coefficient, but its origin is qualitatively different. In addition to the distinct spin structure, the leading low-energy contribution arises from the interference between the tree-level $Z^\prime$ exchange and the QED amplitude, rather than from the interference involving a one-loop BSM amplitude.
The sign and size of the $Z^\prime$ coefficient depends on the model parameters and the charges of the initial state quarks ($Q_q$).
For a heavy $Z^\prime$ with universal couplings to light quarks, we expect the process $p p \to Z^\prime$ to be dominated by the valence up quarks, hence we assume $Q_q = 2/3$. This leads to a negative (positive) form factor at small $\hat{s}$ and $g_q g_t >0$ ($g_q g_t <0$).

At the operator level, the leading low-energy contribution can be represented by:
\begin{equation}
    \mathcal{O}_{tDG}
    =
    \left(
    \bar t_R\gamma^\mu T^A t_R
    \right)
    D^\nu G^A_{\nu\mu}.
\label{eq:OtDG}
\end{equation}
For an off-shell gluon with momentum $q^2=\hat{s}$, this operator produces precisely the momentum dependence proportional to $\hat{s}/M^2$ appearing in Eqs.~\eqref{eq:EFTscalar} and \eqref{eq:EFTvlf}. Using the gluon equation of motion,
\begin{equation}
    D^\nu G^A_{\nu\mu}
    =
    g_s
    \sum_q
    \bar q\gamma_\mu T^A q,
\end{equation}
$\mathcal{O}_{tDG}$ is equivalent, up to equations of motion, to color-octet four-fermion interactions of the form
\begin{equation}
    \left(
    \bar t_R\gamma^\mu T^A t_R
    \right)
    \left(
    \bar q\gamma_\mu T^A q
    \right). \label{eq:4fermionOp}
\end{equation}
Therefore, sufficiently below the BSM threshold the Scalar and VLF models reduce to the same effective operator structure, but with different Wilson coefficients. 
The same applies to the $Z^\prime$ scenario, which can also be described (at small $\hat{s}$) by the operator in Eq.~\eqref{eq:4fermionOp}.
Although information about the spin of the heavy particles is consequently retained in the EFT matching coefficient, it only affects the normalization of the dimension 6 Wilson coefficient for the EFT operators at low $\hat{s}$.

Fig.~\ref{fig:FormFactors_real} shows the behavior of the three form factors defined in 
Eqs.\eqref{eq:fmuZero2}, \eqref{eq:fmuZero2VLF} and \eqref{eq:ZprimeFF}.
Since the overall normalization of the curves is not relevant, we rescale $F_{Z^\prime}$ in order to make a comparison between the form factor shapes more direct.
Also, since the VLF and Scalar form factors are maximum around $\hat{s} \sim 4 M^2$, while $F_{Z^\prime}$ peaks at $\hat{s} = m_{Z^\prime}^2$, we take $m_{Z^\prime} = 2M$.
For the $Z^\prime$ model we set $Q_q=2/3$, $g_q=-1.2\times10^{-2}$ and $g_t=0.35$, which results in a positive interference term at low $\hat{s}$. 
The left plot shows the dependence on $\hat{s}$ in the low energy ($\hat{s} \lesssim M$) regime, where the EFT description is valid.
As expected from the above discussion, the VLF form factor grows faster than the Scalar one. The slope of the $Z^\prime$ form factor depends on the choice of parameters and for the values assumed, it displays a slope similar to the Scalar model.

\subsubsection*{Form Factors above Threshold}
\begin{figure}
    \centering
\includegraphics[width=0.49\textwidth]{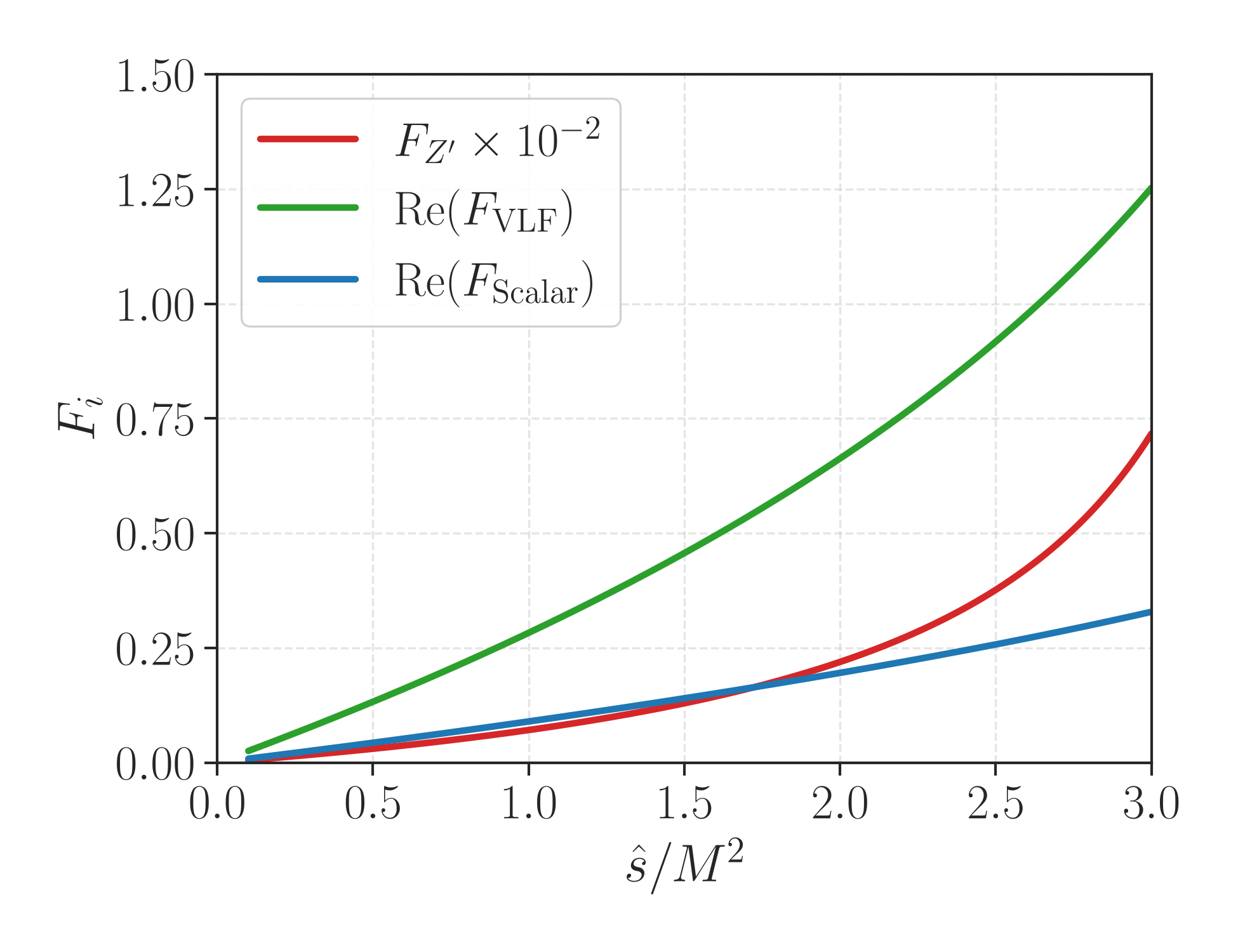}
    \includegraphics[width=0.49\textwidth]{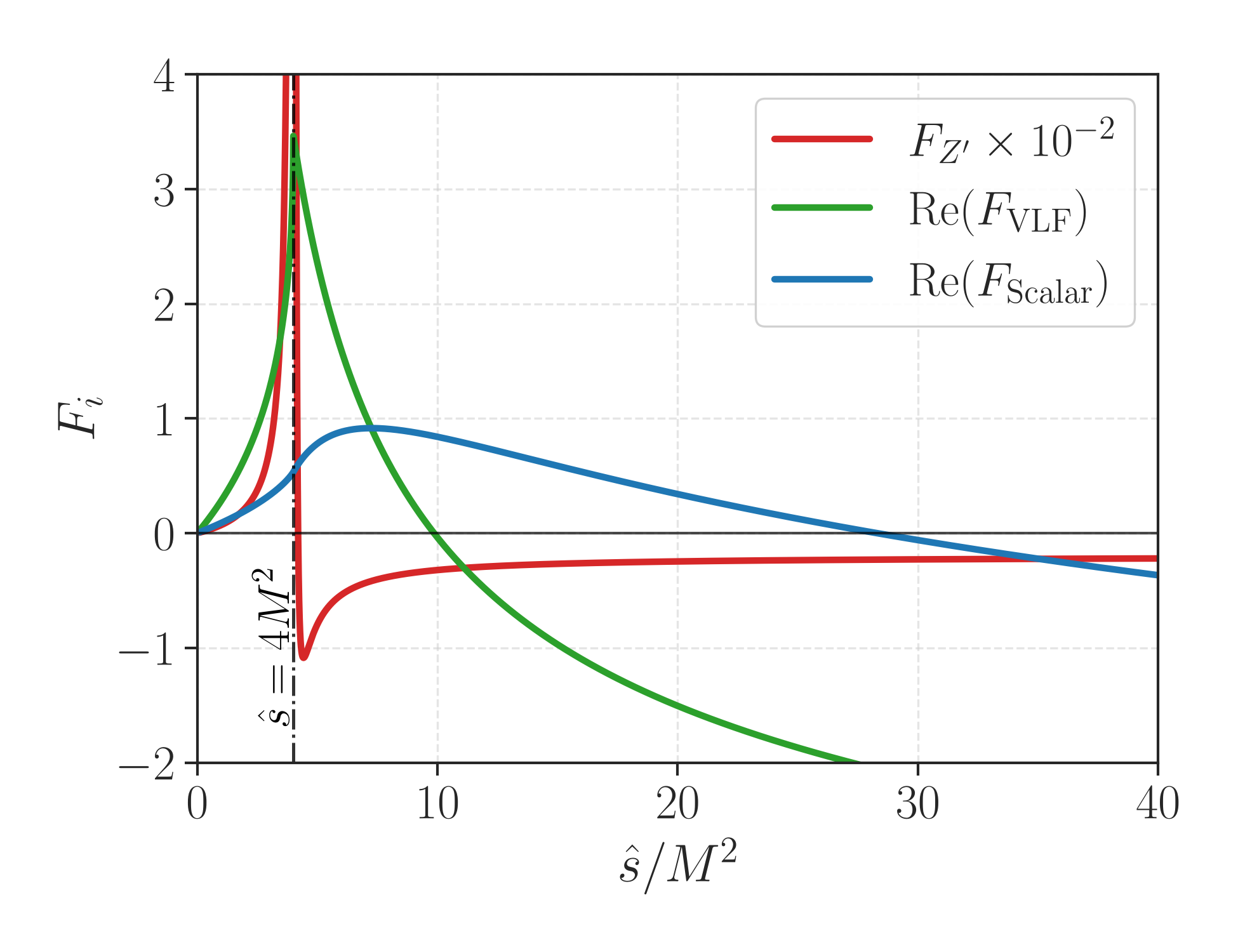}
    \caption{Real part of the dimensionless form factors $F_{\mathrm{Scalar}}$, $F_{\mathrm{VLF}}$, and $F_{Z^\prime}$ as a function of $\hat{s}/M^2$.
    The left plot shows the behavior at low $\hat{s}/ M^2$, while the right plot shows the form factors for a wide range of $\hat{s}/M^2$ values.
    The Scalar and VLF form factors are evaluated in the $m_t\to0$ and $m\simeq M$ limits, where $M$ denotes the top partner mass. For the $Z^\prime$ model, the mass and width are chosen as $m_{Z^\prime}=2M$ and $\Gamma_{Z^\prime}=0.01m_{Z^\prime}$, respectively, and the other parameters are chosen as $Q_q=2/3$, $g_q=-1.2\times10^{-2}$, and $g_t=0.35$ (see text for details).}
    \label{fig:FormFactors_real}
\end{figure}

As the partonic energy approaches the threshold, $\hat{s}/M^2 \simeq 1$, 
the growth is no longer linear in $\hat{s}$ and the local expansion ceases to be adequate.
The behavior at large $\hat{s}$ is shown in Fig.~\ref{fig:FormFactors_real}b, where $F_{\mathrm{Scalar}}$ exhibits a broad resonant behavior around $\hat{s}/M^2 \simeq 7$. In addition, for $\hat{s}/M^2 \gtrsim 28$ the real part becomes negative, leading to a negative interference contribution.
For the VLF model we also observe a resonant structure and a negative interference at large $\hat{s}$. In this case, however, the resonance is around  $\hat{s}/M^2 \simeq 4$ and is significantly sharper than in the Scalar model. Moreover, $\mathrm{Re}(F_{\mathrm{VLF}})$ becomes negative at smaller $\hat{s}$ values ($\hat{s}/M^2 \gtrsim 10$) when compared to the Scalar scenario. 
Finally, the $Z^\prime$ model exhibits the typical narrow resonant behavior around $\hat{s} \simeq 4 M^2 = m_{Z^\prime}^2$ and the form factor becomes negative for $\hat{s} > m_{Z^\prime}^2$.

From these results it becomes clear that exchanging the spin assignments of the BSM particles affects both the shape of the resonant structure and the energy scale at which the interference term becomes negative.
Although the results shown in Fig.\ref{fig:FormFactors_real} only apply to the process $q \bar{q} \to t\bar{t}$ and within the approximations $m_t, \Delta M \simeq 0$, the form factor behavior leaves a similar footprint on experimentally measurable observables. 
We will explore these effects in more detail in the next section,
where we investigate how these features manifest themselves in the differential $t\bar{t}$ distributions obtained from Monte Carlo simulations and including the leading BSM 1-loop corrections.

\subsection{Differential distributions}
\label{sec:differential}

In this section we analyze the distributions for the invariant mass distribution of $t\bar{t}$, $m(t\bar{t})$, the top transverse momentum ($p_T$) and the emission angle in the $t\bar{t}$ center of mass frame ($\cos\theta^*$).
All the differential distributions discussed in this and the following sections were obtained from parton-level Monte Carlo simulations generated with \texttt{MadGraph5}~v3.6.3~\cite{Alwall:2011uj} at 13.6~TeV center of mass energy. In order to efficiently populate the high-energy tails of the distributions, an invariant-mass bias was applied during event generation. All samples were generated using the \texttt{NNPDF23\_nnlo\_as\_0119\_qed} parton distribution function (PDF) set~\cite{Ball:2012cx}. The SM background and the $Z^\prime$ signal predictions were computed at leading order (LO), while the predictions for the Scalar and VLF models include the leading one-loop BSM contributions for the interference terms (see Eq.~\eqref{eq:amp}).

Since our main goal is to assess the LHC potential to distinguish between the three spin possibilities, we consider three benchmark models that, as much as possible, yield similar contributions to top-quark observables.
Table~\ref{tab:benchmarks} lists the parameters adopted for each benchmark model. As discussed above, choosing the $Z^\prime$ mass to be twice the top-partner mass leads to an enhancement in the top-quark distributions around the same energy scale. Furthermore, the coupling values in each model lead to comparable deviations from the SM around the resonant region.
Table~\ref{tab:benchmarks} also includes a broad $Z^\prime$ benchmark with $\Gamma_{Z^\prime} = 0.2 m_{Z^\prime}$. This benchmark will be used in Sec.~\ref{sec:shape_discrimination} when testing the distinguishability of the benchmarks. However, for simplicity, in this section we will show results only for the narrow width case.
A sample of 1M SM events was generated at LO, while for the Scalar and VLF models, between $870$k and $2$M events were generated in order to achieve a statistical precision between $1\%$ and $5\%$ per bin across most of the relevant invariant mass range. For the $Z^\prime$ model, a dedicated sample of 2M events was produced.

\begin{table}[htpb]
    \centering
        \begin{tabular}{|ccc|}
        \hline
        \textbf{Model} & Masses  & Couplings\\
        \hline
        \textbf{SM} &  $m_{t} = 172$~GeV & --\\
        \hline
        \textbf{VLF} &  $(m_{\psi},m_{\phi}) = (1.5,1.4)$~TeV & $y_{\rm DM} = 3.8$\\
        \textbf{Scalar} & $(m_{\varphi},m_{\chi}) = (1.5,1.4)$~TeV & $y_{\rm DM} = 7.5$\\
        $\mathbf{Z^\prime}$ ($\Gamma_{Z^\prime}/m_{Z^\prime} = 1\%$)  & $m_{Z^\prime} = 3.0$ TeV & $(g_q,g_t) = (-0.012,0.35)$\\
        $\mathbf{Z^\prime}$ ($\Gamma_{Z^\prime}/m_{Z^\prime} = 20\%$) & $m_{Z^\prime} = 3.0$ TeV & $(g_q,g_t) = (-0.035,1.58)$\\
        \hline
    \end{tabular}%
    \caption{Parameters for the benchmark models used in this work.}
    \label{tab:benchmarks}
\end{table}

\begin{figure}[t]
    \centering
        \includegraphics[width=0.49\linewidth]{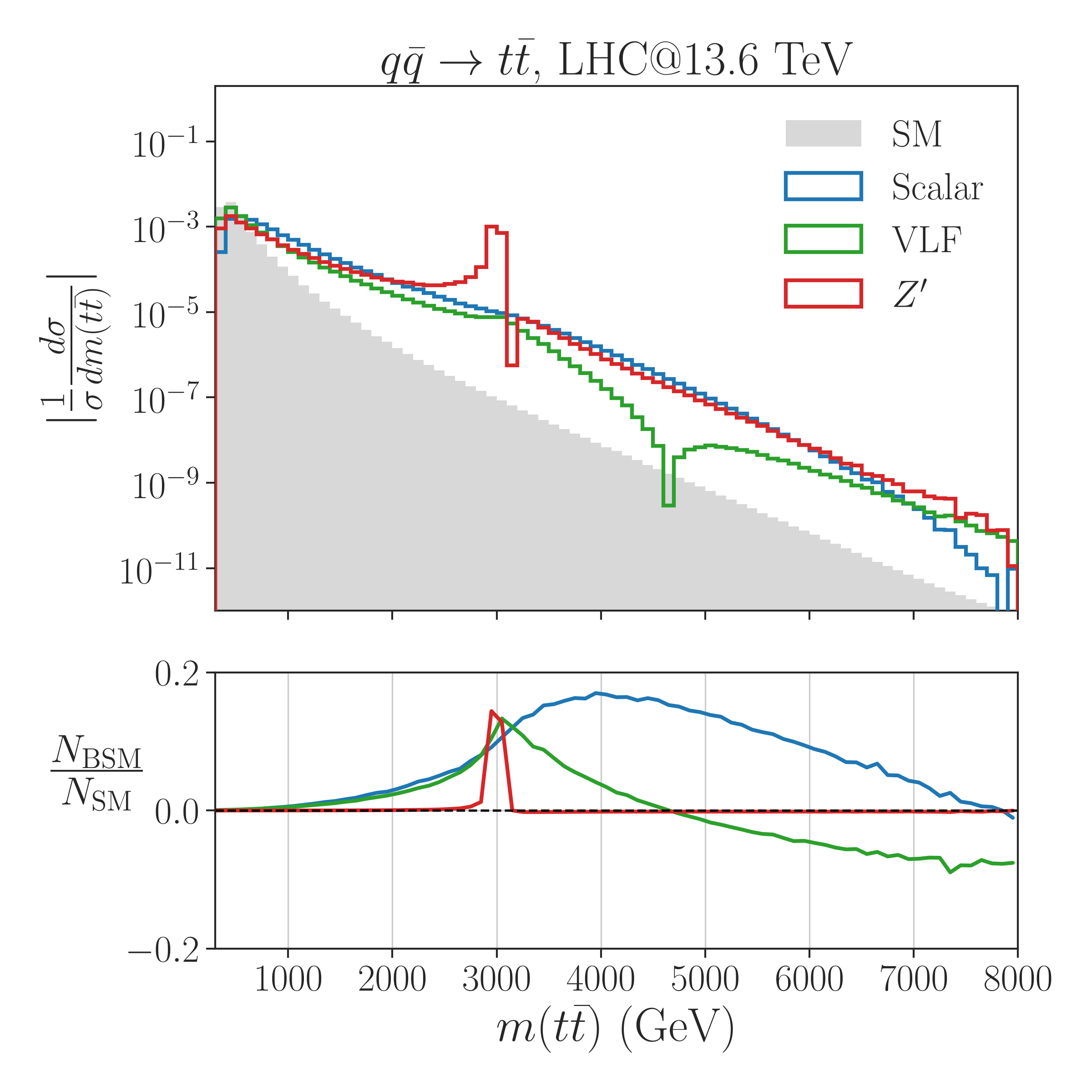}
        \includegraphics[width=0.49\linewidth]{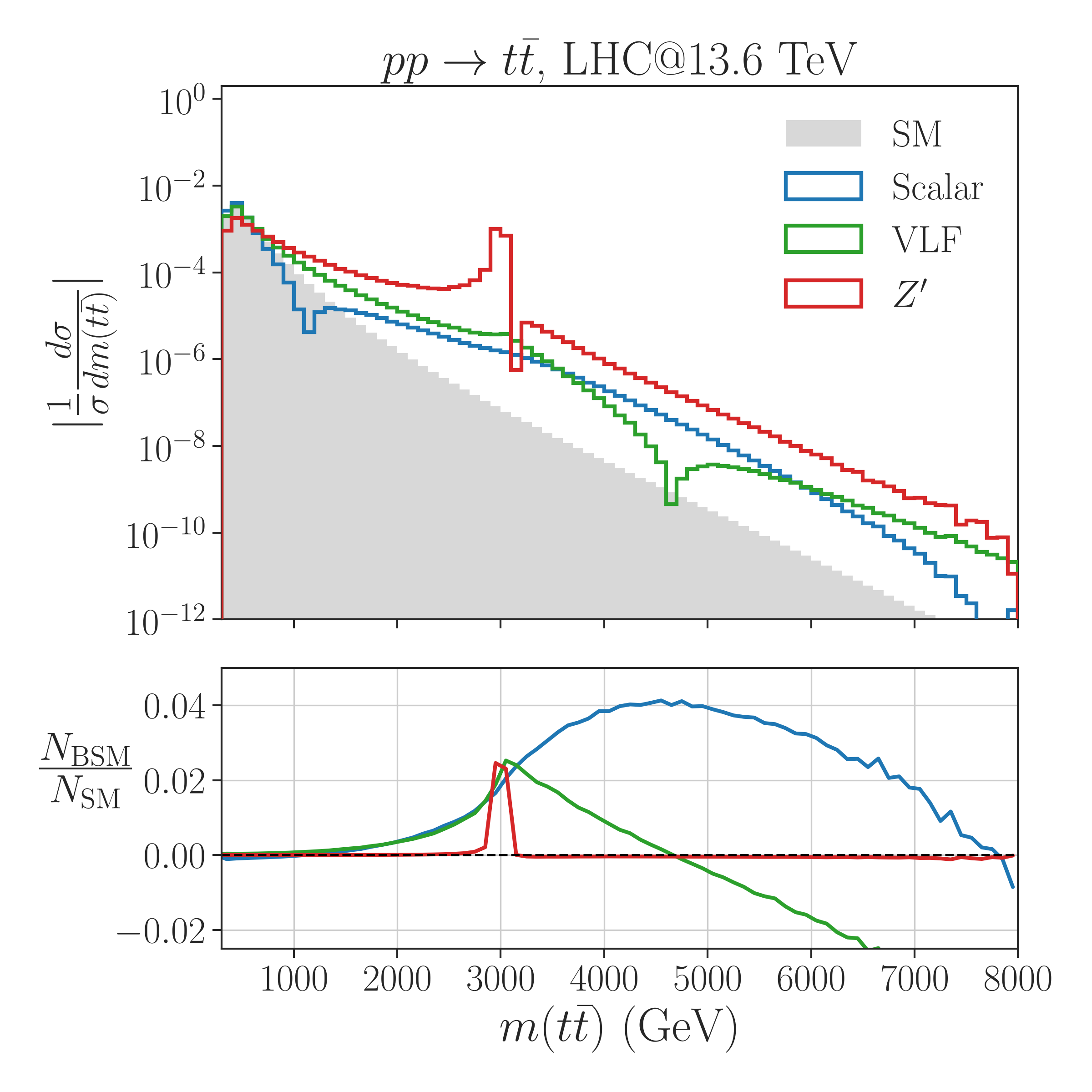}    
    \caption{Normalized invariant mass distributions of the $t\bar{t}$ system. 
    The filled histogram represents the SM leading-order Born contribution. 
    The blue and green histograms correspond to the BSM interference terms 
    for the VLF and Scalar models, respectively, while the red histogram 
    shows the pure BSM squared amplitude for the $Z^\prime$ model. 
    The distributions are evaluated for the
    model parameters defined in Table~\ref{tab:benchmarks}, with $(M, m) = (1.5, 1.4)~\mathrm{TeV}$ and $m_{Z^\prime} = 3$~TeV.
    Since the loop-induced 
    interference contributions can be negative, the upper panels display absolute values. 
    The lower panels show the ratio of each BSM prediction to the SM expectation.}
    \label{fig:mttdists_combined}
\end{figure}

 As a first step, we study the $m(t\bar{t})$ distribution for the subprocess $q\bar{q}\to t\bar{t}$.
 Since at parton level the invariant mass is equal to the partonic center of mass energy, $m(t\bar{t}) = \sqrt{\hat{s}}$, the distribution is directly proportional to the form factors discussed in Sec.~\ref{sec:formFactors}. Hence the $m(t\bar{t})$ distributions allows us to validate our analytical results.
 Fig.~\ref{fig:mttdists_combined}a shows the normalized distributions for the three benchmark models as well as the SM (filled histogram).
 The BSM interference terms defined in Eq.~(\ref{eq:amp}) are shown by the blue and green histograms for the Scalar and VLF models, respectively. For the $Z^\prime$ model, the leading BSM contribution, consisting of the pure squared amplitude together with its interference with QED, as defined in Eq.~(\ref{eq:amp_Zp}), is shown by the  red histogram. Since interference effects can lead to negative contributions, the upper panel shows the absolute value of the differential distributions, while the lower panel displays the ratio of each BSM contribution to the corresponding SM prediction.

The distributions confirm the behavior of the form factors discussed in Sec.~\ref{sec:formFactors}  and shown in Fig.~\ref{fig:FormFactors_real}. In particular, when compared to the SM distribution, the VLF model exhibits a broad resonant-like enhancement around $m(t\bar{t}) \simeq 2 M = 3$~TeV, followed by a negative interference contribution appearing near $m(t\bar{t}) \simeq 3.1 M = 4.7$~TeV. The Scalar model also presents a resonant-like enhancement, although it is broader than in the VLF case and peaks at  $m(t\bar{t}) \simeq 2.7M = 4$~TeV. The negative interference in the Scalar model arises at higher invariant masses, beginning at approximately $m(t\bar{t}) \simeq 5.3 M = 8$~TeV. 
For the $Z^\prime$ model, the numerical results are also consistent with the analytical behavior of the form factor in the $Z^\prime$ model. In particular, we observe a pronounced resonance around $m(t\bar{t})\simeq m_{Z^\prime}$, as expected from the dominance of the Breit-Wigner contribution in this region. At higher invariant masses, $m(t\bar{t})\gtrsim m_{Z^\prime}$, the contribution is small, but negative as a result of the interference between the BSM $Z^\prime$ exchange and the SM Born contribution from QED.

The analysis of the subprocess $qq \to t\bar{t}$ serves primarily as a tool for understanding how the BSM effects arise and how the spin assignment influences the structure of the distributions in a simplified setup. We now turn to the full process $pp \to t\bar{t}$, which also includes contributions from the gluon-fusion channel $gg \to t\bar{t}$.
Despite depending on more complex loop functions, this channel still preserves many of the features discussed above. 
However, for the Scalar and VLF models, a purely analytical analysis of the interference terms in the full process becomes considerably more complex.
For this reason, we rely on Monte Carlo simulations to investigate how the BSM effects manifest themselves in the invariant mass distribution of the top–antitop pair produced in proton–proton collisions.

Fig.~\ref{fig:mttdists_combined}b shows the normalized invariant-mass distributions  for  $pp \to t\bar{t}$ at parton level, where we adopt the same parameter choices used for the $q \bar{q} \to t \bar{t}$ subprocess.  
Although many of the features discussed previously remain present, one notable difference appears in the Scalar model: for $m(t\bar{t}) \lesssim 0.7M = 1.1$~TeV, the interference term is negative, due to the gluon-fusion contribution. At higher mass values it becomes positive and reaches its maximum around $m(t\bar{t}) \simeq 3 M = 4.5$~TeV, turning negative again at even higher invariant masses. 
The VLF model presents a behavior similar to the one observed in the quark-initiated process, except for the fact that the negative interference at high $m(t\bar{t})$ becomes larger (in absolute value).
In this way, the effects associated with the spin assignment become even more evident in the invariant mass distribution of the full process.
We also point out that while for the SM the $gg \to t \bar{t}$ cross-section is $\sim 5$ times larger than the $q \bar{q} \to t \bar{t}$ cross-section, the same is not true for the BSM contributions. While for the Scalar and VLF models both cross-sections are similar, the $Z^\prime$ model only contributes to the quark-initiated process.
This leads to the significant reduction in the BSM/SM ratio shown in the lower panel of Fig.~\ref{fig:mttdists_combined}b.

The results for the Scalar and VLF models shown in Fig.~\ref{fig:mttdists_combined} assume a DM mass ($m$) nearly degenerate with the top partner mass $M$.
While the position of the resonant-like enhancement in both models is primarily determined by $M$, the energy scale at which the negative interference region starts does depend on $m$.
However, since the degenerate scenario is harder to be probed by direct searches (see Sec.~\ref{sec:directSearches}), this is the region where $t\bar{t}$ measurements can be competitive~\cite{Lessa_2024}. Therefore in the following we will always consider benchmarks with $m \sim M$.

\begin{figure}[t]
    \centering
        \includegraphics[width=0.49\linewidth]{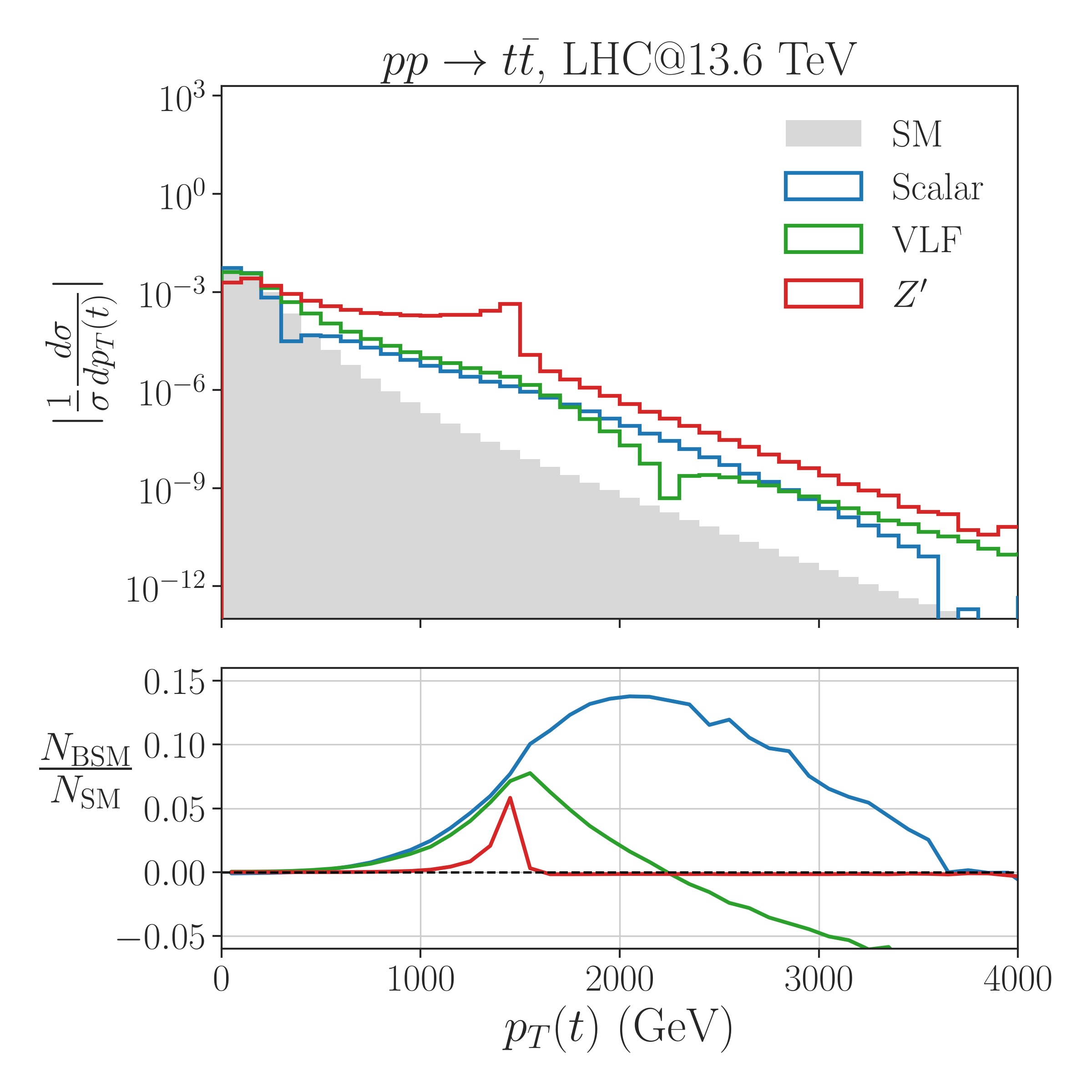}
        \includegraphics[width=0.49\linewidth]{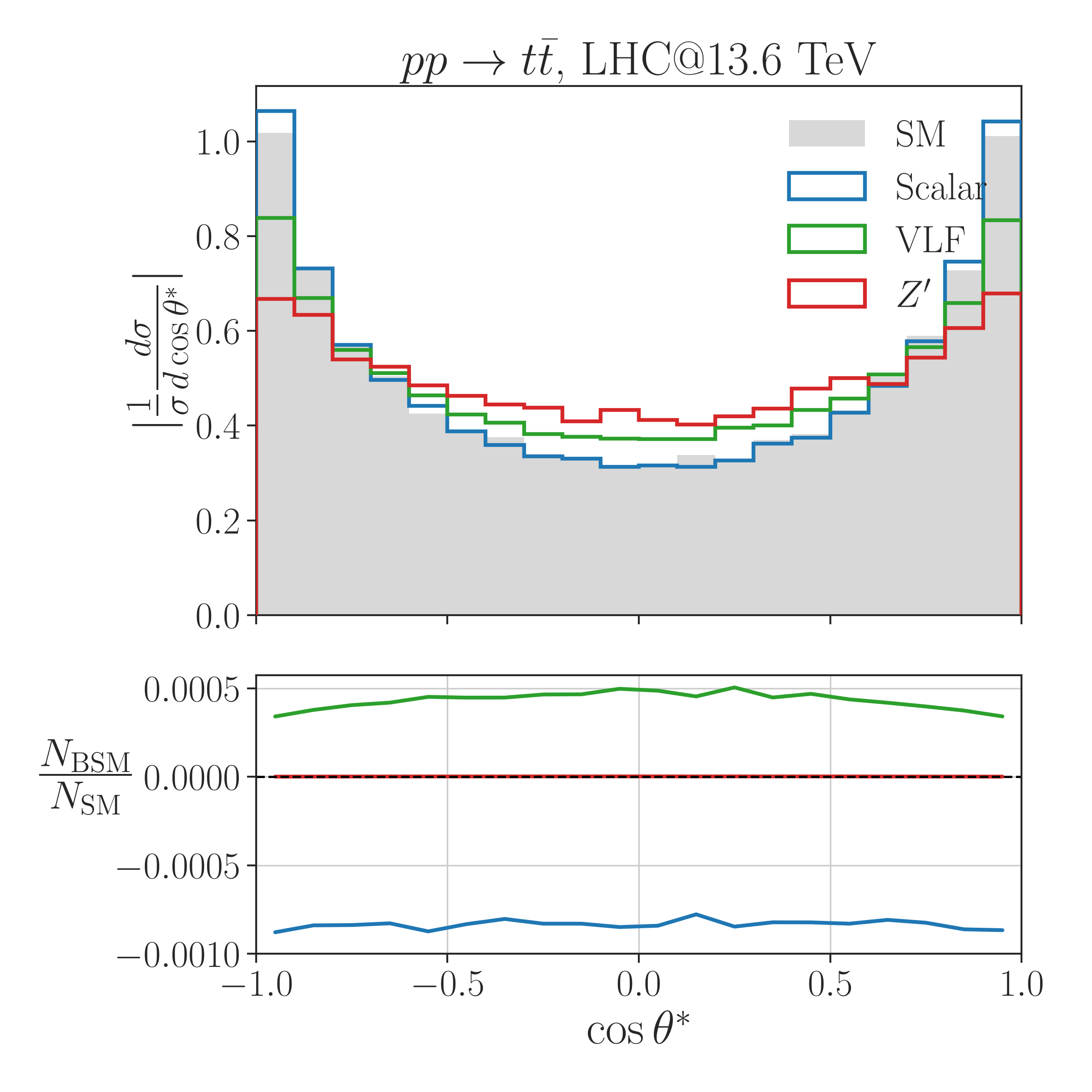}   
    \caption{Transverse momentum and $\cos{\theta^*}$ distributions of the top quark for the SM, Scalar, VLF and $Z^\prime$ models.
    The filled histogram represents the SM leading-order Born contribution. 
    The solid blue and green histograms correspond to the BSM interference terms 
    for the VLF and Scalar models, respectively, while the solid red histogram 
    shows the pure BSM squared amplitude for the $Z^\prime$ model. 
    The model parameters are the same used in Fig.~\ref{fig:mttdists_combined}.
    Since the loop-induced 
    interference contributions can be negative, the upper panels display absolute values. 
    The lower panels show the ratio of each BSM prediction to the SM expectation.}    
    \label{fig:kdists_combined}
\end{figure}

It is also interesting to consider the transverse momentum and $\cos\theta^*$ distributions for each model. Fig.~\ref{fig:kdists_combined} shows these distributions for the same choice of model parameters considered in Fig.~\ref{fig:mttdists_combined}.
As we can see, the transverse momentum, which is symmetric under $t \leftrightarrow \bar{t}$ at parton level, displays a behavior similar to the invariant mass distribution. The main difference is that the maximum of the distributions (when compared to the SM) take place at $p_T(t) \simeq M$ ($1.5 M$) for the VLF (Scalar) model.
Furthermore, around the maximum, the ratio with respect to the SM background is considerably larger than for the $m(t\bar{t})$ distribution.
We also see that the angular distribution in the $t \bar{t}$ center of mass frame has a similar behavior for all models, except for the overall normalization of the signal.
The $\cos\theta^*$ distribution for the $Z^\prime$ model displays a behavior similar to the VLF case, however, due to its small total cross-section, it is not visible in the lower plot.
Since the $p_T$ leads to the largest excess with respect to the SM\footnote{Although all of our results are at parton level, we have verified that the inclusion of showering only impacts the low $p_T$ bins, which are not relevant for our discussion.}, in Sec.~\ref{sec:shape_discrimination} we will exploit this distribution to quantify the potential of the High Luminosity (HL) LHC for differentiating between these benchmarks.

\section{Direct Search Constraints}\label{sec:directSearches}

\begin{figure}
\begin{minipage}[b]{0.32\textwidth}
    \centering
    \includegraphics[scale = 0.95]{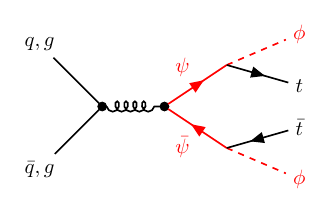}  \\
    (a) VLF
\end{minipage}
\hspace{3pt}
\begin{minipage}[b]{0.32\textwidth}
    \centering
    \includegraphics[scale = 0.95]{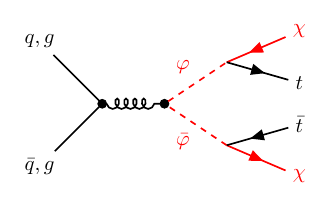} \\
    (b) Scalar
\end{minipage}
\hspace{3pt}
\begin{minipage}[b]{0.32\textwidth}
    \centering
    \includegraphics[scale = 0.95]{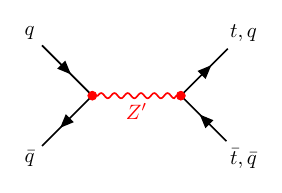}  \\
        (c) $Z^\prime$
   \end{minipage}
   \caption{Diagrams for the on-shell production and decay of top partners in the VLF (a) and Scalar (b) models. The on-shell production of $Z^\prime$ and its decay to top or light quarks is shown in (c).}
    \label{fig:onShell}
\end{figure}

In Section~\ref{sec:observables} we have shown how the $t\bar{t}$ distributions are affected by the BSM contributions from distinct spin scenarios.
As we have seen, the main difference between the Scalar, VLF and $Z^\prime$ models appear above the threshold production, $\hat{s} \gtrsim M^2$, where $M$ corresponds to the BSM mass scale.
However, at these energies, the new particles can be produced on-shell, as shown in Fig.~\ref{fig:onShell}, and tested by LHC searches.
In this Section we give a brief overview of the relevant constraints from direct searches. As we will show, the benchmarks listed in Table~\ref{tab:benchmarks} are beyond the current sensitivity of these searches.

The constraints on the VLF and Scalar models can be obtained through a re-interpretation of searches for top partners decaying to top and dark matter. In particular, searches for pair production of stops with direct decays to neutralinos and top provide the main bounds on these scenarios.
Note that these constraints are independent of $y_{\rm DM}$, since it only affects the top partner width and, for the values considered in this work, the decays are always prompt.
The $Z^\prime$ model, on the other hand, is constrained by searches for resonances in the $q\bar{q}$ and $t\bar{t}$ final states.
We have used \textsc{SModelS}~v3.1.1~\cite{Altakach:2024jwk,MahdiAltakach:2023bdn,Alguero:2021dig,Kraml:2013mwa} to obtain the constraints for each point in the parameter space of the VLF, Scalar and $Z^\prime$ models. The production cross-sections and branching ratios were computed at LO using \textsc{MadGraph5}~v3.6.3.

\begin{figure}
    \centering
    \includegraphics[width=0.49\linewidth]{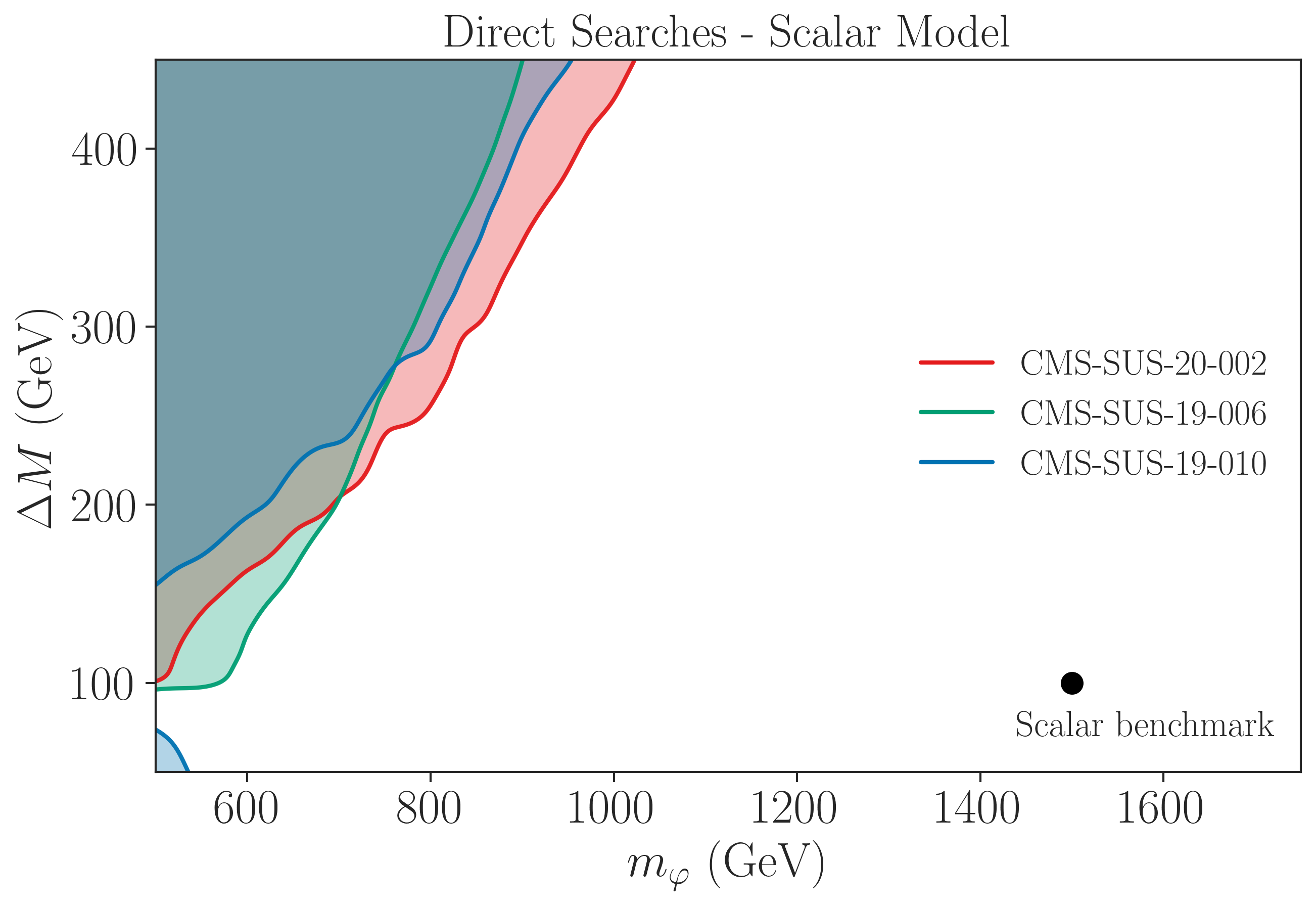}
    \includegraphics[width=0.49\linewidth]{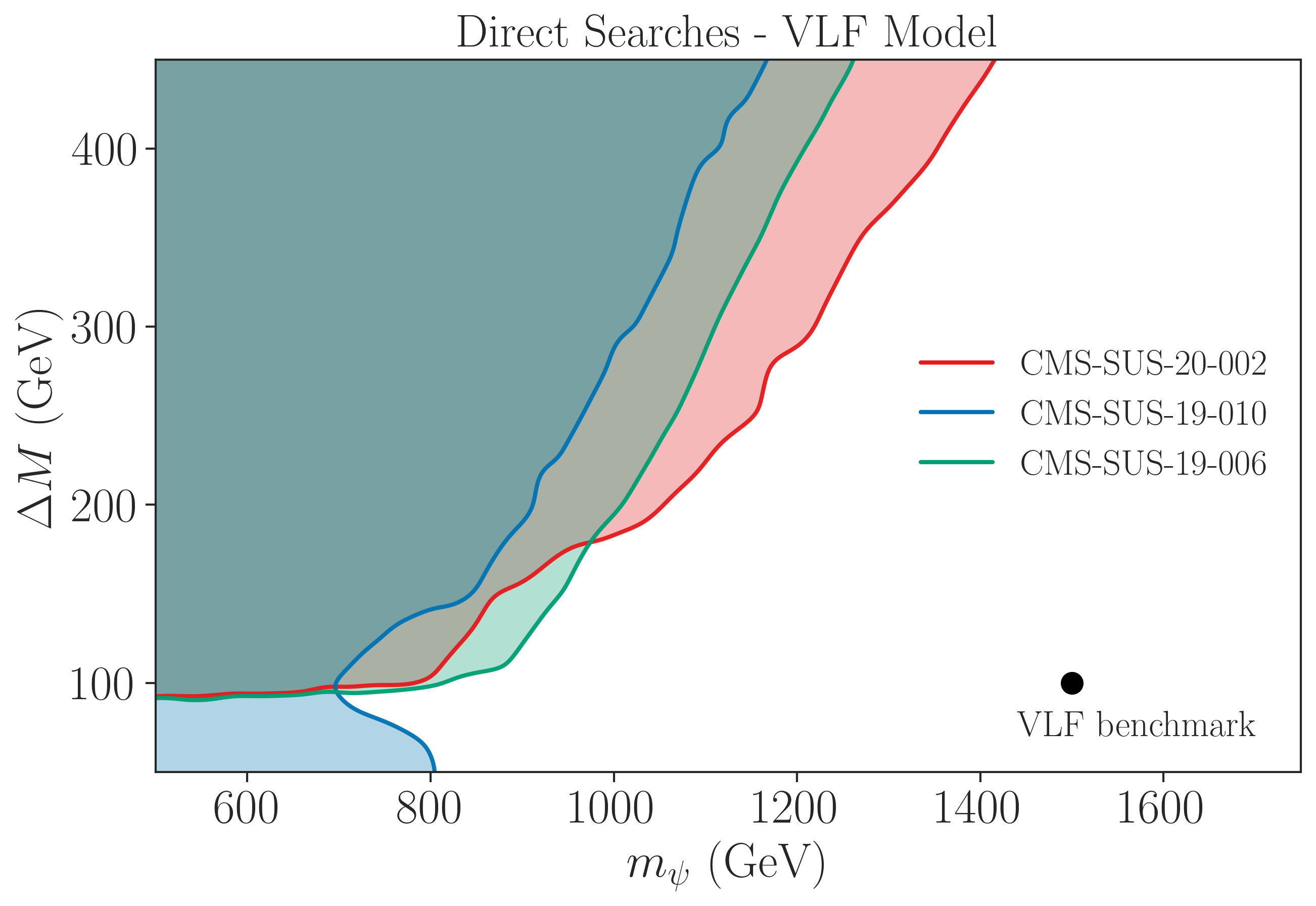}
    
    \caption{Regions excluded at 95\% C.L. by Run 2 LHC searches for top partners decaying into a top quark and Dark Matter. The left (right) plot shows the constraints for the Scalar (VLF) model. The blue, red and green curves display the leading exclusions from CMS-SUS-19-010~\cite{CMS-SUS-19-010}, CMS-SUS-20-002~\cite{CMS-SUS-20-002} and CMS-SUS-19-006~\cite{CMS-SUS-19-006}, respectively. The benchmark points defined in Table~\ref{tab:benchmarks} are shown by the filled dots. See text for more details.}
    \label{fig:directSearches}
\end{figure}

The constraints on the Scalar  model is shown in Fig.~\ref{fig:directSearches}a as a function of the top partner mass ($m_\varphi$) and the mass difference $\Delta M = m_\varphi - m_\chi$. For $\Delta M > m_t$, where the top partners can decay to on-shell tops,
the leading constraints are from the combined CMS search~\cite{CMS-SUS-20-002} for stops in the 0, 1 or 2 leptons channels (red curve). Scalar top partners are excluded up to $m_\varphi \simeq 1.1$~TeV for $\Delta M < 500$~GeV.
Once $\Delta M \lesssim m_t$, searches for decays to off-shell tops become more sensitive. In this compressed region, the CMS search~\cite{CMS-SUS-19-006} for jets and missing $E_T$ (MET) is the most constraining one (green curve). Finally, for $\Delta M \lesssim 50$~GeV, the CMS search from Ref.\cite{CMS-SUS-19-010}, which targets compressed stops (blue curve)  excludes masses up to $m_{\varphi} \simeq 550$~GeV. The benchmark point considered for the Scalar model, with masses $(m_{\varphi},m_{\chi})=(1.5,1.4)$~TeV is shown by the black point. It falls into the compressed region and, due to its small cross-section, is beyond the sensitivity of current searches.

Fig.~\ref{fig:directSearches}b shows the corresponding exclusion curves for the VLF model. The results are very similar to the Scalar scenario, except for the larger production cross-sections, which lead to a larger excluded region. In particular, for the large $\Delta M$ region the exclusion extends  up to $m_\psi \simeq 1.4$~TeV for $\Delta M \simeq 500$~GeV.
In the off-shell region, $\Delta M < m_t$, the MET and compressed stop searches once again become more constraining, excluding masses up to $m_\psi \simeq 0.8$~TeV.
The VLF benchmark from Table~\ref{tab:benchmarks} is shown by the black point in the right plane, being far from the current sensitivity.

\begin{figure}
    \centering
    \includegraphics[width=0.49\linewidth]{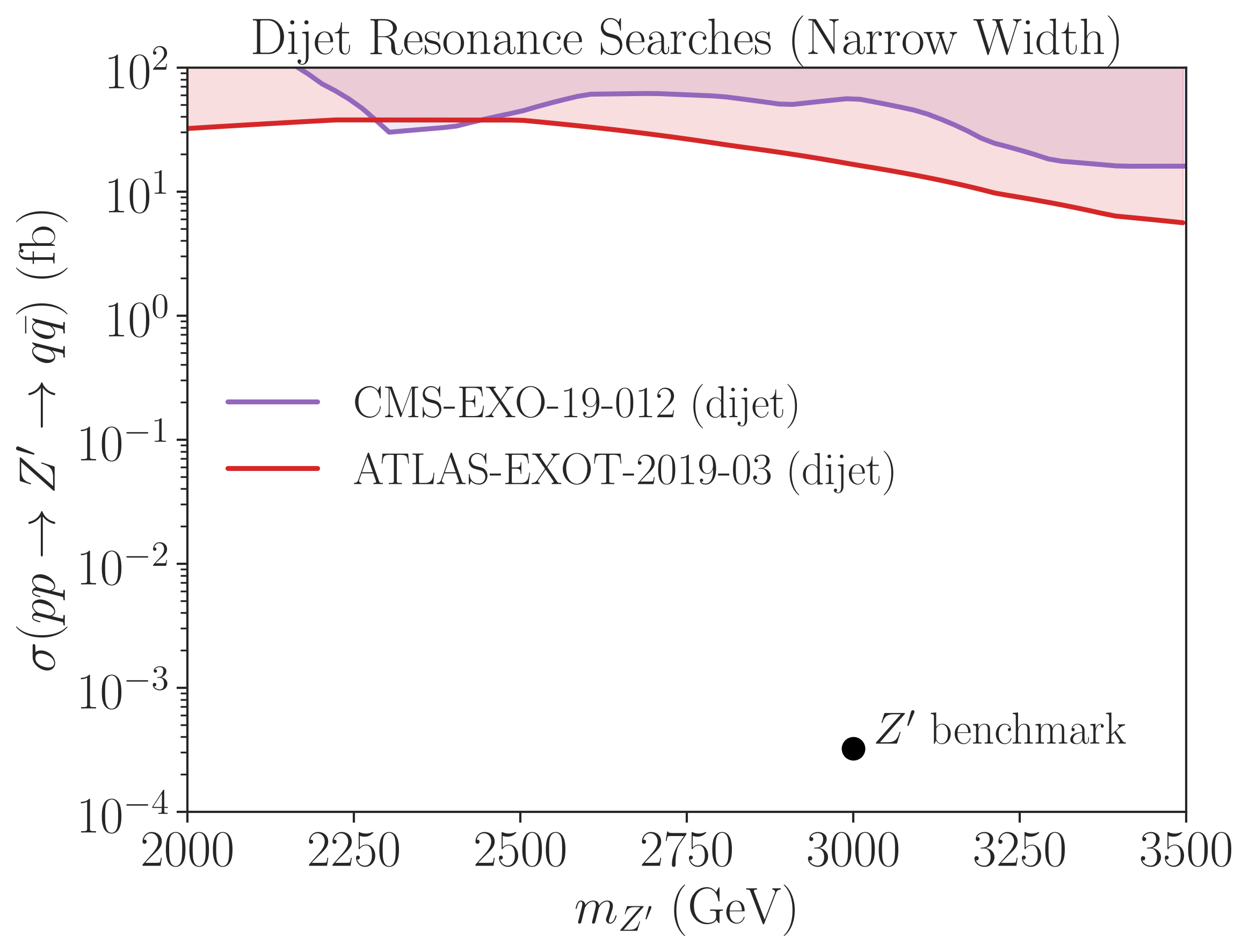}
    \includegraphics[width=0.49\linewidth]{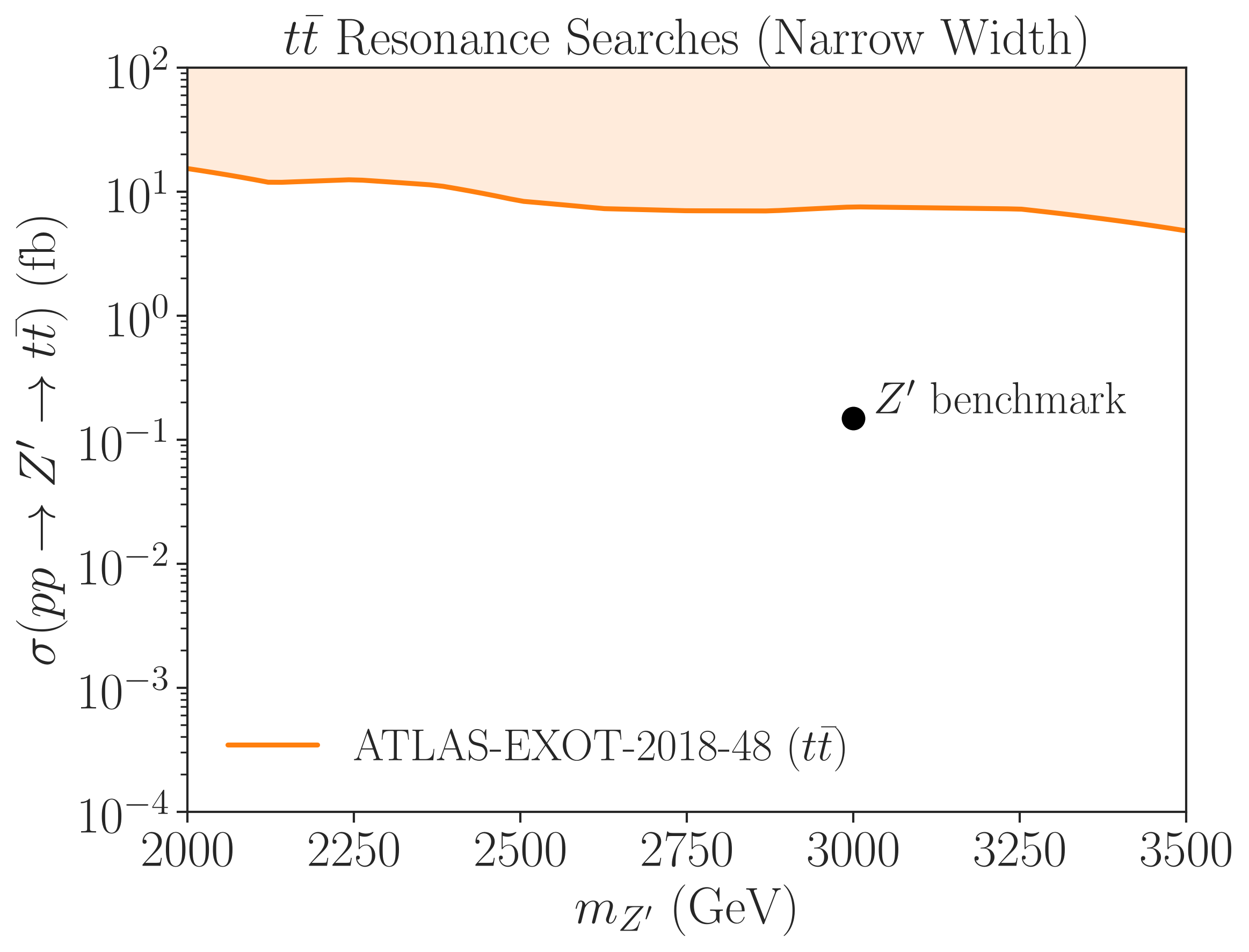}
    \caption{Constraints on the a narrow $Z^\prime$ resonance decaying to light jets (left panel) and a pair of top quarks (right panel). The benchmark point defined in Table~\ref{tab:benchmarks} is shown by the filled dots. See text for more details.}
    \label{fig:directSearchesZp}
\end{figure}

Finally, in Fig.~\ref{fig:directSearchesZp} we show the constraints for the $Z^\prime$ model for the two relevant decay channels: $Z^\prime \to q\bar{q}$ (a) and  $Z^\prime \to t\bar{t}$ (b). 
The results shown are for a narrow resonance with $\Gamma_{Z^\prime}/m_{Z^\prime} < 1\%$.
For the $q \bar{q}$ final state (Fig.~\ref{fig:directSearchesZp}a), the ATLAS search for di-jet resonances from Ref.\cite{ATLAS-EXOT-2019-03} presents results up to $m_{Z^\prime} = 3.5$~TeV and is the most constraining, except for a small region around around $m_{Z^\prime} \simeq 2.3$~TeV.
For the high mass region the CMS di-jet search~\cite{CMS-EXO-19-012} becomes the dominant one.
The benchmark from Table~\ref{tab:benchmarks} has $BR(Z^\prime \to q \bar{q}) = 0.2\%$, being well beyond the sensitivity of di-jet searches. However, since $BR(Z^\prime \to t \bar{t}) = 99.8\%$, searches for resonances decaying to a pair of tops are the most relevant ones, as shown in Fig.~\ref{fig:directSearchesZp}b. In this case the ATLAS search from Ref.\cite{ATLAS-EXOT-2018-48} is the most sensitive, excluding cross-sections larger than a few fb. The benchmark considered, however, has $\sigma(p p \to Z^\prime) BR(Z^\prime \to t \bar{t})= 0.15$~fb being beyond the current sensitivity.
The constraints for the broad width benchmark from Table~\ref{tab:benchmarks} with $\Gamma_{Z^\prime}/m_{Z^\prime} = 20\%$ are similar to the narrow width case, hence we do not show them in Fig.~\ref{fig:directSearchesZp}. 
For this benchmark, however, we have $\sigma(p p \to Z^\prime \to q\bar{q})  =2.5\times 10^{-3}\, \rm{fb}$ and $\sigma(p p \to Z^\prime \to t \bar{t})  = 2.4\, \rm{fb}$. Despite being closer to the current sensitivity, this benchmark is not excluded by the Run 2 searches.

\section{Shape discrimination in the \texorpdfstring{$p_T$}{pT} spectrum}
\label{sec:shape_discrimination}

In Section~\ref{sec:observables}, we discussed how the spin assignments of new states affect the $t\bar{t}$ kinematic distributions. Although the Scalar, VLF, and $Z^\prime$ models give rise to distinct kinematic features, it remains unclear whether these differences can be exploited to distinguish the underlying scenarios in the presence of the large SM background.
This Section addresses this question using the four benchmark scenarios listed in Table~\ref{tab:benchmarks}, including two $Z^\prime$ scenarios corresponding to narrow and broad resonances. As discussed previously, these benchmarks lie beyond the sensitivity of LHC Run 2 searches.
However, the HL-LHC will dramatically increase the accumulated luminosity, allowing to explore these scenarios.
Therefore a natural next step is to investigate 
whether these distinct kinematic features can be exploited to quantitatively discriminate between the signal hypotheses.
To this end, we assume that deviations from the SM predictions have been observed in the data and that these deviations originate from one of the benchmark scenarios in Table~\ref{tab:benchmarks}. We then investigate whether the observed deviations can help identify the underlying BSM scenario and, in particular, constrain the spin of the new state.
As discussed in Sec.~\ref{sec:differential}, the transverse momentum of the hardest top quark appears to be the most promising observable for identifying an excess over the SM prediction. We therefore focus on a binned likelihood analysis of the $p_T$ spectrum to assess the extent to which this observable can discriminate between the benchmark hypotheses.
Finally, we will ignore the effects of systematical uncertainties, since estimating these for the HL-LHC is beyond the scope of the current work.

Our analysis relies on the following steps:
\begin{enumerate}
    \item First we assume one of the benchmark scenarios from Table~\ref{tab:benchmarks} to be the truth and use it to generate synthetic data, including the pure (LO) SM contribution.
    \item We then perform a simple $\chi^2$ fit for each of the spin hypothesis considered (Scalar, VLF or $Z^\prime$). This allows us to determine the parameters (masses and couplings) which best describe the synthetic data within each scenario.
    \item Finally, using the best fit parameters, we compute a likelihood ratio test to determine whether specific spin hypothesis can be excluded given the synthetic data.
\end{enumerate}
The above steps are repeated using the Scalar, VLF and the broad $Z^\prime$ benchmarks as the truth hypothesis.
When determining the best fit parameters for the Scalar and VLF scenarios we generate MC events for several values of the top partner mass, while fixing $\Delta M = 100$~GeV.\footnote{For the synthetic data considered, the specific choice of dark matter mass has no significant impact on the analysis.} For the $Z^\prime$ model, we vary $m_{Z^\prime}$ in steps of 100~GeV and generate a sample of MC events for each value and for the narrow ($\Gamma_{Z^\prime}/m_{Z^\prime}=0.01$) and broad ($\Gamma_{Z^\prime}/m_{Z^\prime}=0.2$) resonance cases.
For convenience the best fit is performed on the relevant mass ($M$) and the signal strength $\theta$, where $\theta = y_{\rm DM}^2$ for the Scalar and VLF models and  $\theta = g_q g_t$  for $Z^\prime$ model.

\subsection{Statistical Method}
\label{subsec:stats_method}

The best fit parameters $(\hat{\theta},\hat{M})$ to the synthetic dataset are obtained by maximizing the binned Poisson likelihood~\cite{Cowan1998, Barlow1989, Cowan2011}. For a binned distribution, the likelihood for a given hypothesis $H$ is:
\begin{align}
\mathcal{L}_{H}(\theta,M)
=
\prod_k
\frac{
\left[n_k^{H}(\theta,M)\right]^{N_k}
e^{-n_k^{H}(\theta,M)}
}{
N_k!
},\label{eq:poissoonLLHD}
\end{align}
where $N_k$ denotes the observed number of events in bin $k$ for a given synthetic dataset and $n_k^{H}(\theta,M)$ is the corresponding expected yield for the model hypothesis $H$:
\begin{align}
n_k^{H}(\theta,M)
=
b_k+S_k^{H}(\theta,M),
\qquad
H\in\{\mathrm{Scalar},\mathrm{VLF},Z^{\prime}\}.
\end{align}
The SM contribution is denoted by $b_k$, while $S_k^{H}$ corresponding to the leading BSM contribution.

For the Scalar and VLF hypotheses, we simply have
\begin{align}
S_k^{H}(\theta,M) = \theta s_k^H(M),
\qquad 
H\in\{\mathrm{Scalar},\mathrm{VLF}\},
\label{eq:expected_yield_scalar_vlf}
\end{align}
with $\theta = y_{\rm DM}^2$.
For the $Z^\prime$ hypothesis both the interference and quadratic BSM terms contribute to the yield:
\begin{align}
S_k^{Z^\prime}(\theta,m_{Z^\prime})
=
\theta^2\,s_{k,\mathrm{sq}}^{Z^\prime}(m_{Z^\prime})
+
\theta\,s_{k,\mathrm{int}}^{Z^\prime}(m_{Z^\prime}),
\qquad
\theta=g_qg_t,
\label{eq:expected_yield_zprime}
\end{align}
where $s_{k,\mathrm{sq}}^{Z^\prime}$ denotes the contribution from the squared $Z^\prime$ amplitude and $s_{k,\mathrm{int}}^{Z^\prime}$ denotes its QED interference. 
Given these definitions, the best fit parameters are simply given by:
\begin{equation}
    \max_{\theta,M} \mathcal{L}_{H}(\theta,M) =  \mathcal{L}_{H}(\hat{\theta},\hat{M})
\end{equation}

Having determined $(\hat{\theta},\hat{M})$ for each hypothesis, we can quantify their compatibility to the synthetic data.
To make use of the asymptotic results for likelihood ratio tests, we define a likelihood $\mathcal{L}(\mu)$ which describes the null hypothesis  ($H_0$) for $\mu=0$ and the true hypothesis  ($H_1$) for $\mu = 1$.
This likelihood is analogous to the one defined in \eqref{eq:poissoonLLHD}, but with total yield defined as \cite{Cox1961, Cox1962, Quandt1974, Algeri2016}:
\begin{align}
n_k(\mu)
=
b_k
+
(1-\mu)\,
S_k^{H_0}(\hat{\theta},\hat{M})
+
\mu\,
S_k^{H_1}(\hat{\theta},\hat{M}).
\label{eq:n_k_nested}
\end{align}
Notice that the best fit parameters $(\hat{\theta},\hat{M})$ above are usually distinct within each hypothesis.

The test statistic is then defined by the likelihood ratio:
\begin{align}
q
=
-2\ln
\left[
\frac{
\mathcal{L}(\mu=0)
}{
\mathcal{L}(\hat{\mu})
}
\right],
\label{eq:test}
\end{align}
where $\hat{\mu}$ denotes the value that maximizes the likelihood.
Since $H_1$ corresponds to the true hypothesis used to generate the synthetic data, we expect $\hat{\mu} = 1$, except for small deviations due to statistical fluctuations in the data.
In the asymptotic limit $q$ is distributed according to a chi-squared distribution with one degree of freedom: $\chi^2(q)$.
The compatibility of the test hypothesis ($H_0$) with the synthetic data can then be quantified by its $p$-value ($p_0$) and the significance $Z$, which is given by:
\begin{align}
Z = \Phi^{-1}(1-p_0) = \sqrt{q_{\rm obs}}.
\end{align}

Finally, to determine the median significance, we employ the Asimov dataset formalism~\cite{Cowan2011}. The Asimov dataset is constructed using a representative data set defined as $N_K^A = n_k^{\rm true} = n_k(\mu=1)$.
Since the Asimov data coincide with the prediction of the true hypothesis, $\mathcal{L}(\mu)$ is maximized  at $\mu=1$, hence $\hat{\mu}=1$. The corresponding Asimov value of the test statistic is therefore:
\begin{align}
q_A
=
2\sum_k
\left[
n_k^{H_0}
-
n_k^{H_1}
+
n_k^{H_1}
\ln
\left(
\frac{n_k^{H_1}}
{n_k^{H_0}}
\right)
\right],
\label{eq:q_test_2}
\end{align}
where $n_k^{H_0} = n_k(\mu=0)$ and $n_k^{H_1} = n_k(\mu=1)$.
The median expected significance is then:
\begin{align}
Z_A
=
\sqrt{q_A}.
\label{eq:Z_asimov}
\end{align}
In particular, the null hypothesis is considered to be excluded at 95\%C.L. if $Z_A > 2$.

In the absence of systematic uncertainties, all expected event yields scale linearly with the integrated luminosity. Consequently we have:
\begin{align}
q_A \propto \mathcal{L} \Rightarrow Z_A \propto \sqrt{\mathcal{L}}.
\end{align}

\subsection{Results}

We employ the framework discussed in Section~\ref{subsec:stats_method} to study whether distinct spin hypothesis can be excluded using the $p_T$ distribution generated by the synthetic data.
We first illustrate the procedure by taking the Scalar model as the true underlying hypothesis and subsequently repeat the analysis for the VLF and $Z^\prime$ hypotheses.

For the Scalar model we use the benchmark in Table~\ref{tab:benchmarks}, which corresponds to the parameters:
\begin{align}
m_{\varphi}=1.5~\mathrm{TeV},\;
m_{\chi}=1.4~\mathrm{TeV}\; \mbox{ and }
y_{\rm DM}=7.5.
\label{eq:ScalarParameters}
\end{align}

Since the SM distribution at low $p_T$ can be significantly impacted by higher order corrections, we restrict our analysis to the range 500~GeV$< p_T < 3500$~GeV and use a uniform binning in steps of 100~GeV:
\begin{align}
p_T\in[0.5, 0.6,0.7,...,3.5]~\mathrm{TeV}.
\end{align}
Within this interval, the significance obtained for the SM as null hypothesis ($H_0$) is approximately $6\sigma$ for an integrated luminosity of $500~\mathrm{fb}^{-1}$.\footnote{We stress that the significances are computed neglecting systematical uncertainties, which will likely reduce their values.}

Using the $p_T$ distribution for the synthetic data generated under the Scalar benchmark hypothesis, we determine the model parameters ($\hat{\theta}$ and $\hat{M}$) which better describe the data under the VLF and $Z^\prime$ hypotheses.
The best fit is computed assuming an integrated luminosity of 500~fb$^{-1}$ and the results are shown in Table~\ref{tab:best_fits}. 
As expected, the best fit parameters for the Scalar model reproduces well the true values.
For the same mass and coupling values the VLF model leads to higher cross-sections and a peak at lower $p_T$ values when compared to the Scalar case, as  discussed in Section~\ref{sec:differential}. Therefore the best fit values for the VLF hypothesis lead to a smaller $y_{\rm DM}$ and a higher $M$.
Finally, for the $Z^\prime$ model the best fit mass is higher than the values for the Scalar and VLF scenarios, in order to reproduce the peak present in the synthetic data.

\begin{figure}[htpb]
    \centering
    \includegraphics[width=0.6\linewidth]{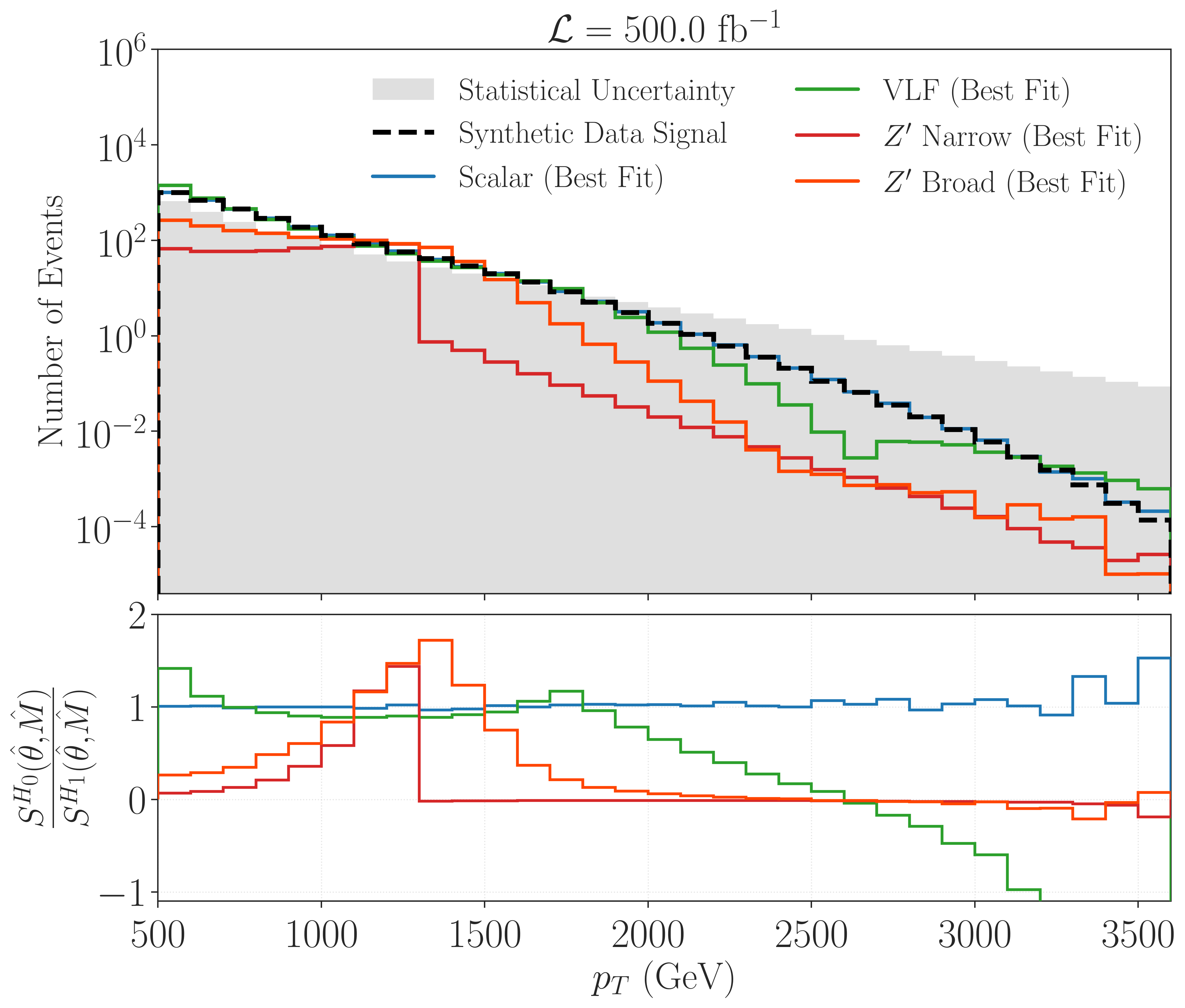}
    \caption{Transverse momentum distribution of the leading top.
    The blue, green, red and orange histograms show the leading BSM contribution for the Scalar, VLF, narrow $Z^\prime$ and broad $Z^\prime$ best fits, respectively.
    The synthetic dataset was generated with the Scalar model as the true hypothesis, while the other models are evaluated at their respective best fit parameter values. The filled band indicate the size of the statistical uncertainty of the synthetic data (SM plus BSM). The lower panel shows the ratio of the BSM contribution from each hypothesis to the corresponding contribution from the Scalar model. Further information on the best fit parameters is given in Table~\ref{tab:best_fits}. }
    \label{fig:best_fit_dists}
\end{figure}

Figure~\ref{fig:best_fit_dists} shows the $p_T$ distribution for the BSM contributions from the best fit models as well as the total statistical uncertainty in each bin (filled histogram).
As we can see, the region with $p_T \gtrsim 1.8$~TeV is dominated by statistical uncertainties and has no significant impact on the determination of the best fit parameters, which are largely determined by the $p_T \lesssim 1.8$~TeV region.
The lower panel shows the ratio between the input signal (Scalar model) and the signal for the best fit models.
As expected, the ratio for the Scalar best fit model is very close to 1, since it corresponds to the correct spin hypothesis.
We also see that the best fit VLF prediction closely reproduces the Scalar contribution over the interval $p_T\in[0.5,1.8]~\mathrm{TeV}$. 
Hence, by adjusting the top partner mass and coupling strength, the VLF model can reproduce much of the broad enhancement generated by the Scalar hypothesis. At higher transverse momenta, however, the VLF contribution decreases more rapidly and eventually becomes negative, while the Scalar contribution remains positive over a wider range. Although this difference becomes increasingly pronounced at large $p_T$, the large statistical uncertainties in this region reduce their impact on the likelihood ratio. Consequently, the VLF hypothesis provides a relatively good description of the Scalar distribution in the region carrying the largest statistical weight.

For the $Z^\prime$ hypotheses, the best fit parameters lead to resonant peaks around $p_T\simeq1.25~\mathrm{TeV}$ and $1.3~\mathrm{TeV}$ for the narrow and broad scenarios, respectively. In both cases, the best fit parameters try to reproduce the synthetic data where it exhibits a significant deviation from the SM. However, the localized resonant structure of the $Z^\prime$ contribution differs significantly from the broader enhancement characteristic of the loop induced Scalar prediction. In particular, the narrow $Z^\prime$ contribution decreases rapidly beyond its resonance, leading to a substantial disagreement with the Scalar distribution at higher transverse momenta. Increasing the resonance width produces a broader enhancement and improves the agreement over a larger range of $p_T$, although differences between the two distributions remain.
From these results it becomes clear that it is significantly easier to distinguish the  $Z^\prime$ (spin 1) hypothesis from the Scalar one than differentiating the VLF and Scalar hypotheses.

\begin{figure}[htpb]
\centering
\includegraphics[width=0.6\linewidth]{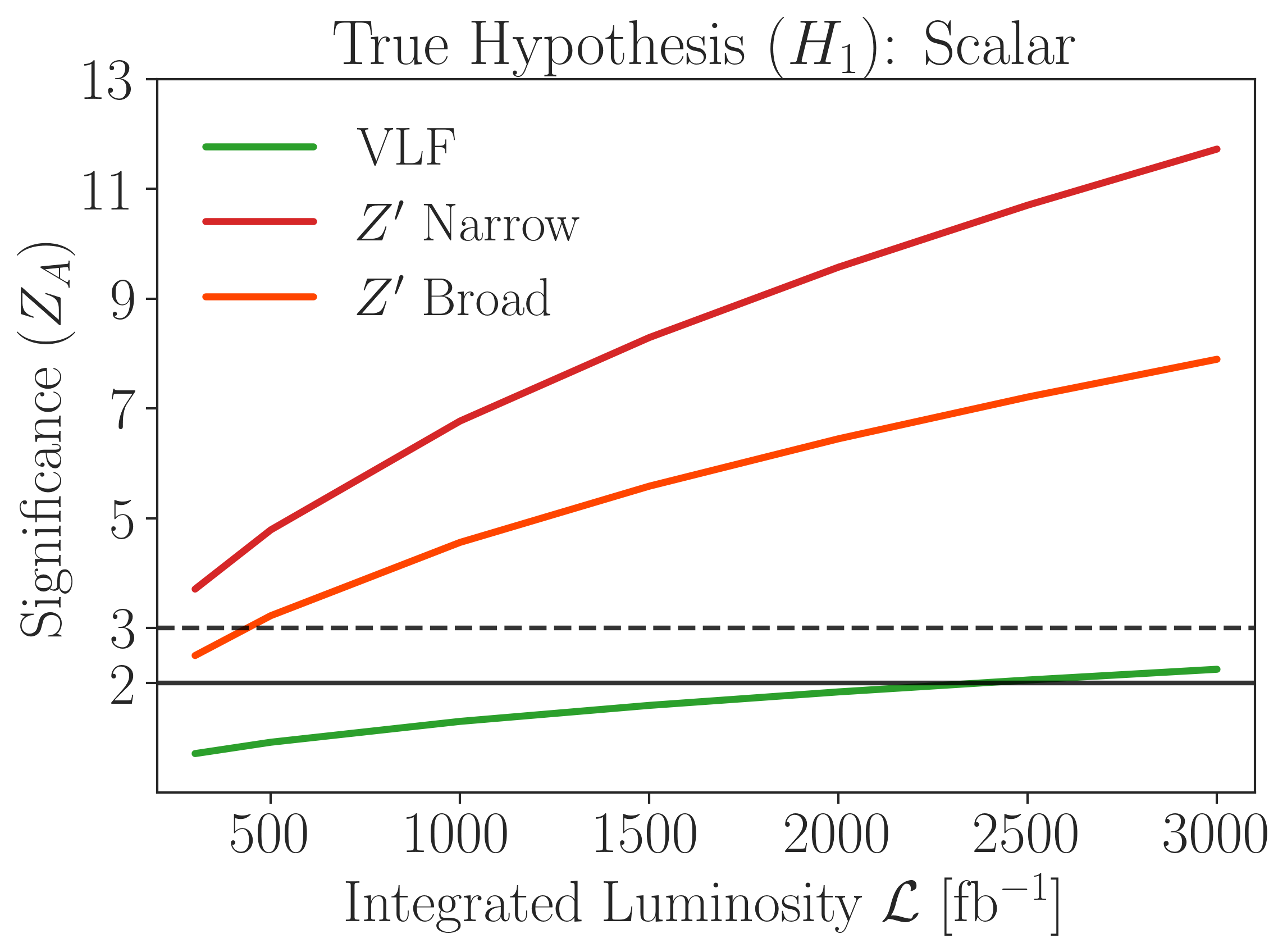}
\caption{Median expected Asimov significance as a function of the integrated luminosity for the shape discrimination analysis. The Asimov dataset was generated assuming the Scalar model with the parameters given in Eq.~\ref{eq:ScalarParameters}. For each test hypothesis, the parameters are fixed to the corresponding best fit values.}
\label{fig:results_Scalar}
\end{figure}

Using the best fit models from Table~\ref{tab:best_fits}, we compute their Asimov significance following the procedure outlined in  Sec.~\ref{subsec:stats_method}.
The results are shown in Fig.~\ref{fig:results_Scalar} as a function of the integrated luminosity $\mathcal{L}$.
At an integrated luminosity of $500~\mathrm{fb}^{-1}$ the narrow $Z^\prime$ hypothesis can be excluded  with a significance of approximately $4.8\sigma$, while the broad $Z^\prime$ scenario reaches $3.1\sigma$. Increasing the luminosity to $3~\mathrm{ab}^{-1}$, corresponding to the target luminosity of the HL-LHC, substantially improves the discrimination power, with both hypotheses exceeding the $5\sigma$ level. The smaller significance obtained for the broad $Z^\prime$ scenario is consistent with its greater compatibility in reproducing the extended enhancement of the Scalar distribution.

\begin{table}[htpb]
    \centering
    \resizebox{\columnwidth}{!}{%
        \begin{tabular}{|ccccc|}
            \hline\hline
            \textbf{Model} & \textbf{Best Fit [GeV]} & \textbf{Coupling} & \textbf{Signal Yield} & $\boldsymbol{\chi^2}$ \\
            \hline
            Scalar ($H_1$) & $1510.2$ & $y_{\rm DM} = 7.6$ & $3032$ & $0.009$ \\
            \hline
            VLF            & $1744.5$ & $y_{\rm DM} = 4.6$ & $3441$ & $0.85$ \\
            
            $Z^\prime$ ($\Gamma_{Z^\prime}/m_{Z^\prime} = 1\%$)  & $2500.1$ & $(g_q,g_t)=(-0.016,\,0.35)$ & $567$  & $23.3$ \\
            
            $Z^\prime$ ($\Gamma_{Z^\prime}/m_{Z^\prime} = 20\%$) & $2882.0$ & $(g_q,g_t)=(-0.030,\,1.58)$ & $1300$ & $10.5$ \\
            \hline\hline
        \end{tabular}%
    }%
    \caption{Best fit parameter values for the different hypotheses, with the Scalar model taken as the true underlying model ($H_1$) and an integrated luminosity of $500~\mathrm{fb}^{-1}$. The coupling column gives $y_{\rm DM}$ for the Scalar and VLF models and the couplings $g_q$ and $ g_t$ for the $Z^\prime$ model. The last two columns show the total number of signal events in the interval $p_T\in[0.5,3.5]~\mathrm{TeV}$ and the best fit $\chi^2$.
}
    \label{tab:best_fits}
\end{table}

As expected from the above discussion, the VLF hypothesis is considerably more challenging to distinguish from the Scalar benchmark. At $500~\mathrm{fb}^{-1}$, the median expected significance is approximately $1\sigma$, increasing to only about $2.2\sigma$ at $3~\mathrm{ab}^{-1}$. This reduced discrimination power can be understood from the 
the fact that the VLF distribution can reproduce well the Scalar model at the low $p_T$ region, as shown in Figs.~\ref{fig:kdists_combined} and \ref{fig:best_fit_dists}.
The main differences between the two models arise at high $p_T$, where the statistical uncertainties are larger.
As a result, only with 2.5~ab$^{-1}$ one expects to exclude the VLF hypothesis at 95\% C.L. ($2\sigma$).

We now repeat the analysis for the scenarios in which the VLF and broad $Z^\prime$ models are taken as the true underlying hypotheses, using the benchmark parameters listed in Table~\ref{tab:benchmarks}. These parameters were chosen such that the corresponding deviations with respect to the SM lead to approximately the same median expected significance as the Scalar benchmark, allowing a more direct comparison of the discrimination power among the different scenarios.
Since the $Z^\prime$ signal leads to a more localized excess, its cross-section is taken to be considerably smaller than the VLF or Scalar scenarios in order to lead to similar significances.

The best fit parameters assuming the VLF as the true hypothesis is shown in Table~\ref{tab:best_fits_VLF}, while the parameters for the broad $Z^\prime$ hypothesis is shown in Table~\ref{tab:best_fits_Zprime}.
As expected, the best fit parameters for the true hypothesis ($H_1$) reproduces well the input values from Table~\ref{tab:benchmarks}. The parameters for the null hypotheses follow the same behavior as before: $m_{Z^\prime} \sim 2 M$ and the Scalar coupling is larger than in the VLF model.

\begin{figure}[htpb]
\centering
\includegraphics[width=0.49\linewidth]{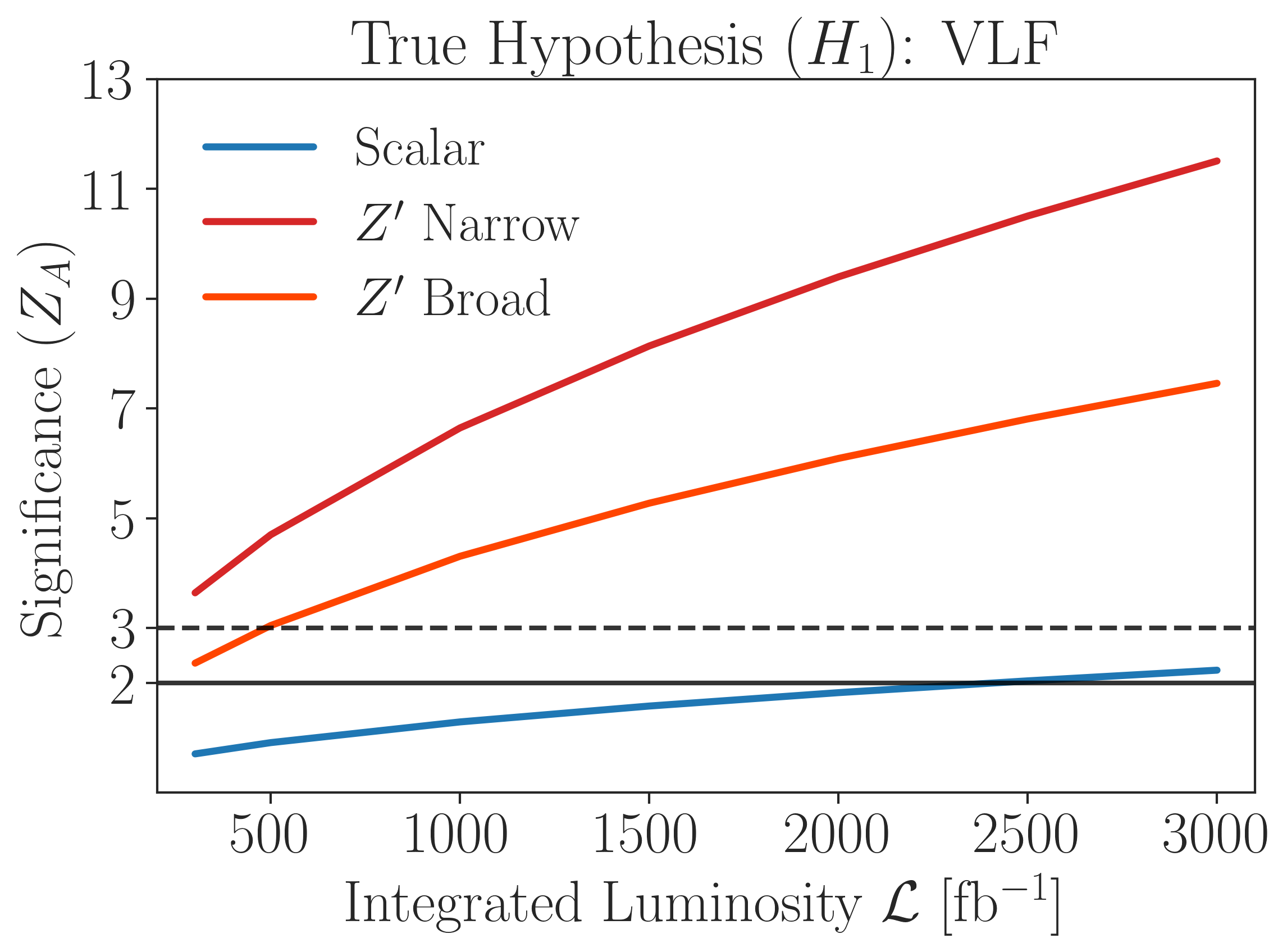}
\includegraphics[width=0.49\linewidth]{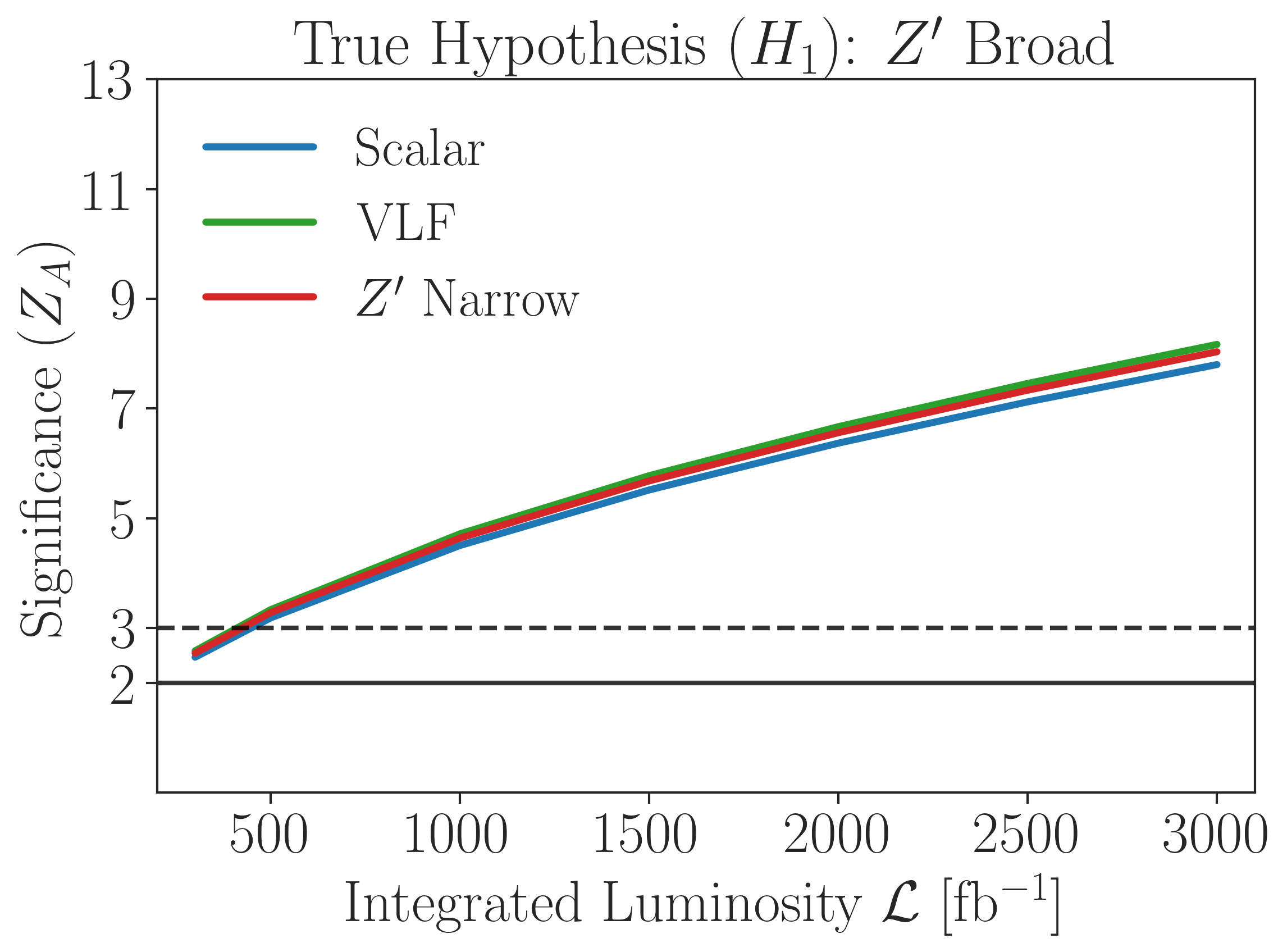}
\caption{Median expected Asimov significance as a function of the integrated luminosity for the shape discrimination analysis. The Asimov dataset in the left and right panels were generated assuming the VLF and $Z^\prime$ models, respectively, with the parameters given in Table~\ref{tab:benchmarks}. For each test hypothesis, the parameters are fixed to the corresponding best fit values.}
\label{fig:results_VLFandZprime}
\end{figure}

The Asimov significances for the VLF and $Z^\prime$ hypotheses are shown in Fig.~\ref{fig:results_VLFandZprime} as function of the integrated luminosity.
The left panel shows the results when the VLF model is taken as the true hypothesis. The significances follow a pattern similar to that observed for the Scalar benchmark. Both $Z^\prime$ hypotheses can be distinguished from the VLF prediction with relatively high significance, exceeding the $5\sigma$ level at $3~\mathrm{ab}^{-1}$. In contrast, the Scalar hypothesis is considerably more challenging to reject, requiring about 2.5~ab$^{-1}$ to exclude it at 95\% C.L.

A different behavior is observed when the broad $Z^\prime$ model is taken as the true hypothesis. As shown in Fig.~\ref{fig:results_VLFandZprime}b, this benchmark exhibits the largest discrimination power among the three scenarios considered. At an integrated luminosity of $500~\mathrm{fb}^{-1}$, all three competing hypotheses can be rejected with significances of approximately $3\sigma$, increasing to around $8\sigma$ at $3~\mathrm{ab}^{-1}$. Although the resulting significances are relatively close, the Scalar model provides the best agreement with the broad $Z^\prime$ prediction, followed by the narrow $Z^\prime$ and VLF hypotheses.
The greater discrimination power obtained for this benchmark can be understood from the different kinematic structures generated by the underlying amplitudes. While the Scalar and VLF models produce broad enhancements associated with loop corrections, the $Z^\prime$ contribution is governed by an $s$-channel propagator, whose resonant structure and interference with the SM lead to a different transverse momentum dependence. Even for a broad resonance, these features cannot be fully reproduced by adjusting the parameters of the loop induced hypotheses. The narrow $Z^\prime$ scenario faces a similar limitation, as its smaller width produces a more localized enhancement that cannot adequately reproduce the extended structure of the broad resonance.

\begin{table}[htpb]
\centering
\resizebox{\columnwidth}{!}{%
        \begin{tabular}{|ccccc|}
            \hline\hline
            \textbf{Model} & \textbf{Best Fit [GeV]} & \textbf{Coupling} & \textbf{Signal Yield} & $\boldsymbol{\chi^2}$ \\
            \hline
            VLF ($H_1$)   & $1505.3$ & $y_{\rm DM} = 3.8$ & $3465$ & $0.008$ \\
            \hline
            Scalar        & $1247.0$ & $y_{\rm DM} = 5.9$ & $3054$ & $0.83$ \\
            $Z^\prime$ ($\Gamma_{Z^\prime}/m_{Z^\prime} = 1\%$)  & $2908.9$ & $(g_q,g_t)=(-0.019,\,0.35)$ & $371$  & $22.2$ \\
            $Z^\prime$ ($\Gamma_{Z^\prime}/m_{Z^\prime} = 20\%$) & $2811.0$ & $(g_q,g_t)=(-0.028,\,1.58)$ & $1306$ & $9.3$ \\
            \hline\hline
        \end{tabular}%
    }
\caption{Best fit parameter values for the different hypotheses, with the VLF model taken as the true underlying model ($H_1$) and an integrated luminosity of $500~\mathrm{fb}^{-1}$. The coupling column gives $y_{\rm DM}$ for the Scalar and VLF models and the couplings $g_q $ and $g_t$ for the $Z^\prime$ model. The last two columns show the total number of signal events in the interval $p_T\in[0.5,3.5]~\mathrm{TeV}$ and the best fit $\chi^2$. }
\label{tab:best_fits_VLF}
\end{table}

\begin{table}[htpb]
\centering
\resizebox{\columnwidth}{!}{%
        \begin{tabular}{|ccccc|}
            \hline\hline
            \textbf{Model} & \textbf{Best Fit [GeV]} & \textbf{Coupling} & \textbf{Signal Yield} & $\boldsymbol{\chi^2}$ \\
            \hline
            $Z^\prime$ ($\Gamma_{Z^\prime}/m_{Z^\prime} = 20\%$) ($H_1$) & $3009.8$ & $(g_q,g_t)=(-0.034,\,1.58)$ & $1372$ & $0.06$ \\
            \hline
            $Z^\prime$ ($\Gamma_{Z^\prime}/m_{Z^\prime} = 1\%$) & $2915.9$ & $(g_q,g_t)=(-0.023,\,0.035)$ & $520$  & $11.0$ \\
            VLF    & $1518.0$ & $y_{\rm DM} = 3.6$ & $2938$ & $11.2$ \\
            Scalar & $1336.1$ & $y_{\rm DM} = 6.0$ & $2670$ & $10.3$ \\
            \hline\hline
        \end{tabular}%
    }
\caption{Best fit parameter values for the different hypotheses, with the $Z^\prime$ model taken as the true underlying model ($H_1$) and an integrated luminosity of $500~\mathrm{fb}^{-1}$. The coupling column gives $y_{\rm DM}$ for the Scalar and VLF models and the couplings $g_q$ and $ g_t$ for the $Z^\prime$ model.  The last two columns show the total number of signal events in the interval $p_T\in[0.5,3.5]~\mathrm{TeV}$ and the best fit $\chi^2$.  }
\label{tab:best_fits_Zprime}
\end{table}

\begin{figure}[htpb]
    \centering
    \includegraphics[width=0.7\linewidth]{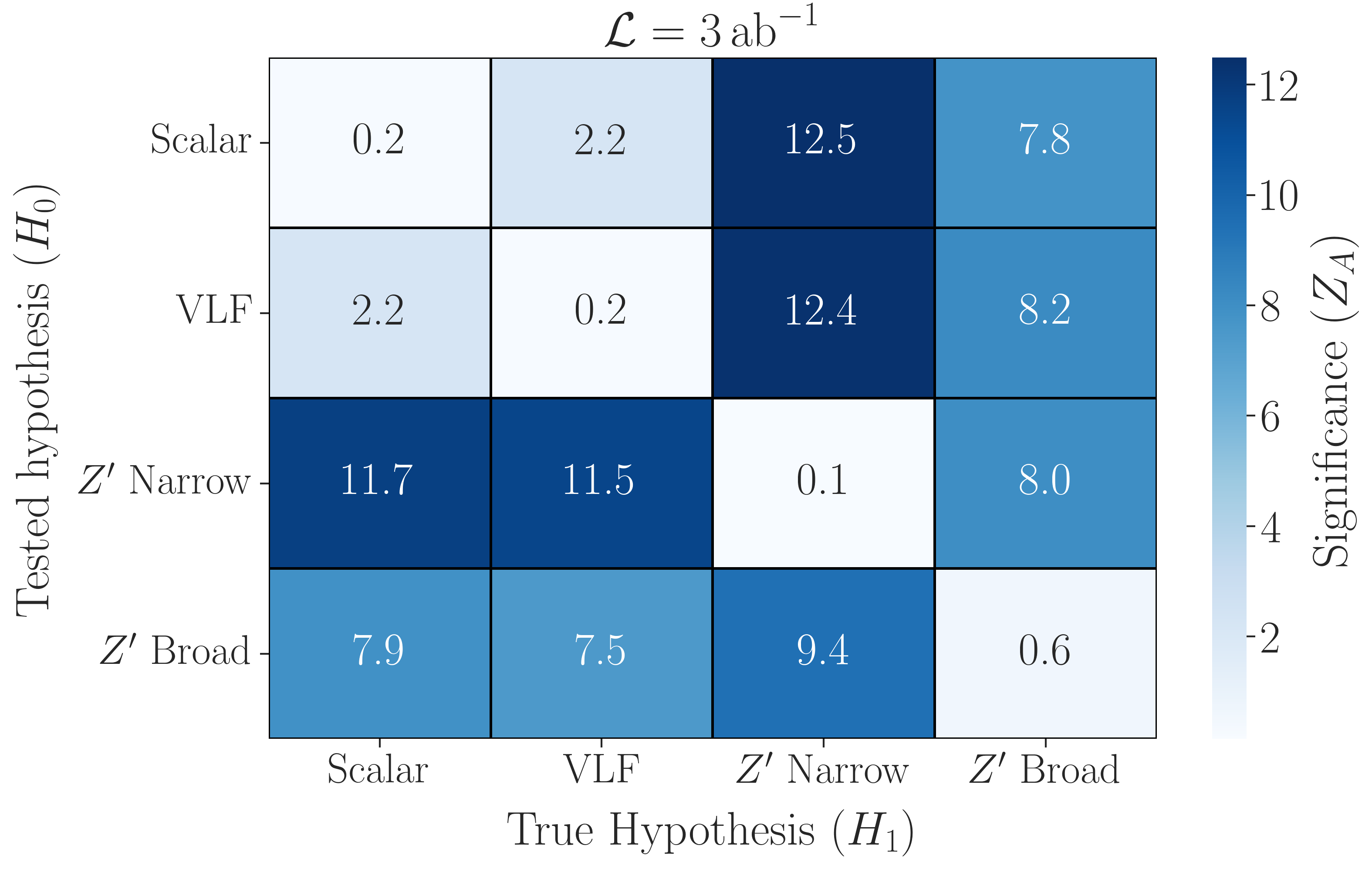}
    \caption{Asimov significances for distinct combinations of the true ($H_1$) and null ($H_0$) hypotheses. The results assume an integrated luminosity of 3~ab$^{-1}$ and neglect systematical uncertainties.\VR{}}
    \label{fig:confMatrix}
\end{figure}

The results obtained under each true hypothesis are summarized in Fig.~\ref{fig:confMatrix}, where we show the Asimov significance for the full HL-LHC integrated luminosity.
Each column represents a given true hypothesis ($H_1$), while the rows correspond to the null hypothesis ($H_0$).
The largest significances are obtained along the $Z^\prime$ rows and column, since this model is clearly distinguishable from the other scenarios.
However, if the synthetic data corresponds to the Scalar or VLF hypothesis, we see that distinguishing these is much more challenging.

\section{Conclusions}\label{sec:conclusions}

The absence of an unambiguous new resonance does not imply that precision measurements can only be used to constrain new physics through effective operators. If new particles lie sufficiently close to the kinematic range probed by an experiment, their virtual effects retain non-local information associated with thresholds, absorptive parts, interference and the spin structure of the particles circulating in the amplitude. In this work we have investigated to what extent this information can be used not only to identify a deviation from the SM, but also to discriminate between different possible microscopic origins of that deviation.

We have used top-pair production as a concrete example and compared three representative top-philic scenarios: a colored scalar top partner coupled to a singlet fermion, a colored vector-like fermion coupled to a singlet scalar, and a spin-one $Z^\prime$ mediator coupling to quarks. The Scalar and VLF scenarios provide a particularly useful comparison because both enter $t\bar t$ production through one-loop corrections and have closely related particle content, while differing in the spin assignments of the particles in the loop. The $Z^\prime$ scenario instead provides a qualitatively different case in which the new contribution is generated at tree level and contains a propagating $s$-channel pole.

The analytic study of the loop-induced top--gluon form factors makes the origin of these differences transparent. The Scalar and VLF amplitudes contain different tensor structures and consequently develop different momentum dependence around the heavy-particle threshold. In the low-energy regime, $s/M^2\ll 1$, these differences reduce to different matching coefficients multiplying the same effective interaction. 
Therefore these models are not distinguishable by their EFT behavior. Near threshold, however, the local expansion ceases to provide an adequate description: the full form factors contain additional information in the position and shape of the enhancement, the onset of imaginary parts, and the pattern of constructive and destructive interference. These features survive in the physical $t\bar t$ invariant-mass and $p_T$ spectra.

We have then asked a more difficult question than whether several fixed benchmark predictions look different. For each competing hypothesis we allow the relevant mass and overall signal strength to adjust so as to provide its best possible description of the assumed signal. The discrimination is therefore driven by differences in shape that cannot be absorbed into a change of the model parameters. This is important for interpreting our results: the analysis is not constructed as a best-case separation of widely different signals, but rather as a test of how much irreducible information remains after the competing models have been allowed to mimic one another.

The distinction between a $Z^\prime$-like contribution and the loop-induced scenarios is comparatively robust. A propagating resonance and its associated interference pattern are difficult to reproduce by changing the parameters of the spin-0 or spin-1/2 models. The more interesting and genuinely difficult problem is the discrimination between the Scalar and VLF hypotheses. Both correspond to loop-induced deformations of the same SM process, both have comparable threshold scales, and changes in the masses and couplings can compensate for part of the difference between their one-dimensional spectra. 
We have shown, however, that given the full HL-LHC luminosity the wrong spin hypothesis can be excluded at 95\% C.L.

The present analysis illustrates the potential of precision measurements to probe the spin of new states. However, this work should not be regarded as exhausting the information available in the process.
First, we have ignored systematical uncertainties, which can play a relevant role and make the model discrimination more challenging.
Second, we have compressed all the kinematical information into a binned one-dimensional transverse momentum distribution. This variable is particularly well motivated because it 
shows the largest deviations with respect to the SM, but it cannot retain all the information contained in the full event. Correlations with the invariant mass and rapidities of the top quarks, angular variables, and additional radiation can respond differently to the tensor and helicity structure of the underlying amplitudes.

This motivates a natural continuation of the present work, where a more complex analysis using the joint phase-space distribution and its correlations could improve the Scalar-VLF comparison.
The results obtained here should therefore be viewed as a deliberately restricted test of how much information can already be extracted from one physically transparent observable, rather than as an upper limit on the achievable discrimination.

Although we have focused on top-quark production, the underlying idea is considerably more general. The essential ingredients are a precisely measured SM process and new particles that contribute virtually over an experimentally accessible range of momentum transfer. Whenever these conditions are satisfied, different ultraviolet completions can produce similar low-energy effective interactions while retaining different non-local structures as the relevant thresholds are approached. Differential measurements can then probe not only the magnitude of a deviation but also its analytic structure.

This situation can arise in other fermionic final states, in di-boson production, in Higgs and electroweak processes, or more generally in any channel in which heavy states enter through vertex, self-energy or box corrections. The most informative observables will depend on the process: invariant masses probe the energy dependence and thresholds of the amplitude, transverse momenta provide related but experimentally distinct projections, while angular and rapidity correlations can retain information about helicity and tensor structures that is absent from an inclusive rate. A multidimensional analysis can combine these complementary pieces of information. In this sense, the strategy explored here is not a top-specific method for spin determination, but an example of a broader program: using precision measurements to move from detecting a deviation, to resolving the momentum dependence of the underlying amplitude, and ultimately to learning about the quantum properties of new particles.

\paragraph{Data availability} 
All data and results presented in this work, together with code required to reproduce the figures, are publicly available on {\sc GitHub}~\cite{githubRepo}. 

\acknowledgments

AL is supported by FAPESP grant no. 2018/25225-9 and 2021/01089-1. The work of VS is supported by the Spanish Government Agencia Estatal de Investigaci\'on Grants No. 
PID2023-148162-C21 and CEX2023-001292-S.

\appendix

\section{One Loop Counterterms}
\label{app:CTs}

As shown in Sec.~\ref{sec:formFactors}, the form factors contributing to the process $qq \to t\bar{t}$ are given by the expressions:
\begin{align}
	\mathcal{F}^\mu_{\mathrm{Scalar}} &= i T^A  \biggl\{ \slashed{p}_1 \left[ p_1^\mu \left( C_1 + 2 C_{11} \right) - p_2^\mu \left( C_1 + 2 C_{12} \right) \right]  \nonumber \\
    &\quad + \slashed{p}_2 \left[ p_2^\mu \left( C_2 + 2 C_{22} \right) - p_1^\mu \left( C_2 + 2 C_{12} \right) \right]  \nonumber \\ 
    &\quad + 2 \gamma^\mu \left(C_{00} + \delta_{R} \right) \biggr\} P_R 
    + 2 i T^A \gamma^\mu \delta_{L} P_L \,,
    \label{eq:fmuOn} \\
    \mathcal{F}_{\mathrm{VLF}}^\mu &= i T^A  \biggl\{ -2 \slashed{p}_1 \left[ p_1^\mu \left( C_1 + C_{11} \right) - p_2^\mu C_{12} \right] \nonumber \\
    &\quad - 2 \slashed{p}_2 \left[ p_2^\mu \left( C_2 + C_{22} \right) - p_1^\mu C_{12} \right] \nonumber \\
    &\quad + \slashed{p}_2 \gamma^\mu \slashed{p}_1 \left( C_0 + C_1 + C_2 \right) \nonumber \\
    &\quad + \gamma^\mu \left[ B_0 + \left( m^2 - M^2 \right) C_0 + 2 p_1^2 C_1 + 2 p_2^2 C_2 - 2 C_{00} + 2 \delta_{R} \right] \biggr\} P_R \nonumber \\
    &\quad + 2 i T^A \gamma^\mu \delta_{L} P_L 
\end{align}
Here, $C_i$ and $C_{ij}$ denote the scalar three-point loop functions, while $\delta_{L,R}$ are the respective counterterms. For completeness we provide analytical expressions for these counterterms below.

The counterterms are computed using \texttt{NLOCT}~\cite{Degrande:2014vpa} within dimensional regularization and imposing on-shell renormalization conditions. For the Scalar model, we obtain\footnote{The expressions in Eq.~(\ref{eq:CTs}) are valid for $M > m + m_t$. Corresponding expressions for the compressed region, $M < m + m_t$, have also been implemented in our numerical analysis.}:
\begin{align}
	\delta_L^{\mathrm{Scalar}} &= \frac{1}{128 \pi ^4 x_t^2 \sqrt{\lambda }} \biggl\{  
    2 \sqrt{\lambda } \left[ x_t \left(x_t+2 x+\log (x)-2\right)-(x-1)^2 \log (x)\right]  \nonumber\\
	&\quad  -4 \left[ x_t \left(x_t+x^2+x-2\right)-(x-1)^3\right] 
    \log \left(\frac{\sqrt{\lambda }-x_t+x+1}{2 \sqrt{x}}\right) \biggr\} \,, \nonumber \\
	\delta_R^{\mathrm{Scalar}} &= \frac{\left(x-1-x_t\right)}{128 \pi ^4 x_t^2 \sqrt{\lambda}} \biggl\{  
    2 \left[ \left(x_t-1\right)^2 + (x-2) x \right] 
    \log\left(\frac{\sqrt{\lambda }-x_t+x+1}{2\sqrt{x}}\right) \nonumber \\
	&\quad + \sqrt{\lambda } \left[ 2x_t-\left(x_t+x-1\right) \log(x)\right] \biggr\}  
    - \frac{1}{64 \pi^4}\left[\frac{1}{\epsilon} + \log\left(\frac{\mu^2}{M^2}\right)\right] 
	\label{eq:CTs}
\end{align}
where $M$ corresponds to the top partner mass, $m$ to the dark matter candidate and we defined:
\begin{align}
    x = \frac{m^2}{M^2} \,, \qquad
    x_t = \frac{m_t^2}{M^2} \,, \qquad
    \lambda = 1 + x^2 + x_t^2 - 2 x x_t - 2 x_t - 2 x 
\end{align}

The divergent term $1/\epsilon$ and the scale-dependent logarithm $\log(\mu^2/M^2)$ appearing in $\delta_R^{\mathrm{Scalar}}$ cancel exactly against the corresponding contributions from $C_{00}$, which is the only divergent loop function entering Eq.~(\ref{eq:fmuOn}).

For the VLF model, the counterterms read:
\begin{align}
    \delta_L^{\mathrm{VLF}} &= -\frac{i}{128 \pi^6 \sqrt{\lambda} y_{\mathrm{DM}}^2 g_s x_t^2} \Bigg\{ 
    \left[x_t \left(x x_t-2 x^2+x+1\right)+(x-1)^3\right] \notag \\
    &\quad \times \left[-2 \log \left(-x_t+\sqrt{x_t^2-2 (x+1) x_t+(x-1)^2}+x+1\right)+\log (x)+\log (4)\right] \notag \\
    &\quad + \sqrt{\lambda} x_t \left(x_t-2 x+2\right) 
    + \sqrt{\lambda} \left((x-1)^2-x x_t\right) \log (x) \Bigg\} \,, \\[1ex]
    \delta_R^{\mathrm{VLF}} &= \frac{1}{128 \pi^4 \sqrt{\lambda} x_t^2} \Bigg\{ 
    \sqrt{\lambda} \left(x_t^2+(x-1)^2\right) \log (x) \notag \\
    &\quad - 2 \left(x_t+x-1\right) \left[ \sqrt{\lambda} x_t+\left(x_t^2-2 x x_t+(x-1)^2\right) 
    \log \left(\frac{\sqrt{\lambda}-x_t+x+1}{2 \sqrt{x}}\right) \right] \Bigg\} \notag \\
    &\quad - \frac{1}{64 \pi^4 }\left[\frac{1}{\epsilon} + \log\left(\frac{\mu^2}{M^2}\right)\right] 
\end{align}

\newpage

\bibliographystyle{unsrt}
\bibliography{references}
\end{document}